\documentclass{aa}  

\usepackage{graphicx}
\usepackage{txfonts}
\begin{document} 

   \title{Revisiting the Local Interstellar Radiation Field using {\em Gaia} DR3}

   \author{
   S.~Bianchi
}

\institute{
INAF-Osservatorio Astrofisico di Arcetri, L. E. Fermi 50, 50125, Firenze, Italy\\
\email{simone.bianchi@inaf.it}
}

\date{}

 
\abstract
{
Dust grains in the interstellar medium are heated by the integrated radiation from stars in the Milky Way. A knowledge of the Local 
Interstellar Radiation Field (LISRF) is thus necessary to interpret observations of dust 
emission in the infrared and constrain (some of) the properties of interstellar grains.
The LISRF representation most widely used in dust modelling still dates back to
the seminal works of \citet{MezgerA&A1982} and \citet{MathisA&A1983}.
}
{
A new version of the LISRF is presented, starting from the photometry of the {\em Gaia} Data Release 3 (DR3) and revisiting available data, among which observations from the 
{\em Pioneer} 10 and 11 probes.
}
{The LISRF contribution by direct starlight is estimated in the {\em Gaia} bands by summing fluxes of all stars in DR3; the LISRF is extrapolated from the optical to the ultraviolet and near-infrared, using the astrophysical parameters provided by DR3 for a subsample of {\em Gaia} stars; the correlation between dust emission at 100 $\mu$m and residual diffuse emission in the {\em Pioneer} and other available maps is exploited to derive the contribution of dust-scattered starlight to the LISRF.
}
{
The new LISRF is significantly redder and emits $\sim$30\% more energy than the old model.
The old LISRF is almost a factor two lower in the near-infrared, while in the optical it 
accounts only for direct starlight. For $|b|<50^\circ$, diffuse starlight contributes on average to $\sim$25\% of the total radiation, $3\times$ more than what predicted by literature estimates at high Galactic latitude.
}
{The new LISRF can modify the predicted mid-infrared dust emission beyond the uncertainties normally assumed between dust models and observational constraints; these differences should be taken into account to redefine the properties of small grains and of the carriers of the mid-infrared emission bands.
}

 
\titlerunning{}
   \maketitle
%

\section{Introduction}

In the Milky Way (MW) and the majority of galaxies, stars emit most of radiation from the ultraviolet (UV) to the near-infrared (NIR). While light at wavelengths $\lambda < 0.0912\,\mu$m is almost entirely suppressed after emission (through ionization of neutral hydrogen), at longer wavelengths the diffuse interstellar medium (ISM) is generally exposed to the integrated radiation from all those stars, nearby and distant, whose light had not been severely extinguished by interstellar dust. 
This interstellar radiation field (ISRF) is generally expressed as an average surface brightness, or as a radiation energy density (the first definition is adopted here).
The ISRF
has a profound impact on the state of interstellar matter, as it regulates the energy states of atoms and molecules and ionizes them and the surface of dust grains, providing free electrons for heating the gas. 
In particular, the ISRF heats dust grains to temperature of a few tens of Kelvin, causing emission at mid and far-infrared (MIR and FIR) and submillimeter (submm) wavelengths \citep[for a general reference, see, e.g.,][]{DraineBook2011}.

While the ISRF is expected to vary within the MW, its 
direct measurement is only possible in the vicinity 
of the  Sun: this so-called local ISRF (LISRF)
can be derived from 
observations of the sky from the
UV to the NIR. 
However, the estimate is complicated: when attempted from the ground, airglow and scattering from the upper atmosphere constitutes an important foreground emission
that needs to be taken into account; observations from near-Earth positions in the outer space reduce (or remove) this, yet another foreground still needs to be considered, zodiacal light (sunlight scattered by interplanetary dust) and its thermal emission (for a review of these and other foregrounds, see \citealt{LeinertA&AS1998}). Finally large-area (ideally, whole-sky) observations are needed, to avoid extrapolations and biases due to specific Galactic regions.
Some of these issues were lifted in the late 70s/early 80s for UV space observations:
those measurements, together with a few constraints in the optical from ground-based observations, were used by \citet*{MezgerA&A1982} to define a model of the LISRF, soon updated by \citet*{MathisA&A1983} to add a further constraint derived from (limited-area) NIR observations in the Galactic plane. 
In the following, I will refer to the LISRF defined in these two works as the MMP model. 

Since 1983, MMP became a popular representation of the LISRF and  \citet{MathisA&A1983} 
was celebrated among other seminal papers in "The first 40 years" issue of this Journal \citep{JonesA&A2009b}. The usage of the MMP formulation was later boosted by the launch of the {\em Herschel} \citep{PilbrattA&A2010} and {\em Planck} \citep{PlanckEarlyI} satellites in 2009,
which significantly increased the availability of MIR, FIR and submm data, for the MW and other galaxies. In fact, starting from the original papers, the MMP has been mainly used as a tool to predict dust emission. For example, models for dust grains in the diffuse MW ISM are set to reproduce the dust emission at high Galactic latitude observed by satellites such as the InfraRed Astronomical Satellite \citep[IRAS;][]{NeugebauerApJL1984}, Cosmic Background Explorer  \citep[COBE;][]{BoggessApJ1992}, and now {\em Herschel} and {\em Planck}, assuming that the heating field is the MMP LISRF \citep[see, e.g.,][]{DesertA&A1990,DwekApJ1997,LiApJ2001,ZubkoApJS2004,CompiegneA&A2011,SiebenmorgenA&A2014,JonesA&A2017}. Also, a (rescaled) MMP model is adopted when computing templates of dust emission under different heating intensities, used to interpret MIR-to-submm observations of galaxies
\citep[see, e.g.,][]{DraineApJ2007,GallianoA&A2011,NersesianA&A2019}.

Only a limited number of new observational constraints has become available since the MMP papers. Most notably, the Diffuse Infrared Background Experiment (DIRBE) aboard COBE has mapped the NIR emission from the whole sky \citep{ArendtApJ1998}: using LISRF estimates from these maps, \citet{DraineBook2011} modified the MMP model, after seeing that its NIR output is lower than observations. On the countrary, \citet{SeonApJS2011} find a lower UV LISRF than predicted by MMP,
using a large sky area survey from the Science and Technology Satellite-1 (STSAT-1). In the
optical range, instead, there has been no major update since the source-counts extrapolations 
and the ground-based estimates used by \citet{MezgerA&A1982}. 
The vast and complete catalogs of optical stellar fluxes from the {\em Gaia} satellite \citep{GaiaDR1mission} now offer the possibility to make a new estimate of the LISRF in the optical.  \citet{MasanaMNRAS2021} use the photometry from {\em Gaia} Data Release 2 \citep[DR2;][]{GaiaDR2summary} complemented with data  from the {\em Hipparcos} satellite \citep{PerrymanA&A1997}, to derive the contribution
of direct starlight to the radiation field; they further exploit a correlation between observed diffuse emission and 100 $\mu$m dust emission from IRAS to estimate the contribution of starlight scattered off dust grains in the ISM (from \citealt{KawaraPASJ2017}; see \citealt{SanoApJ2017} for a review of other results in the literature). However, \citet{MasanaMNRAS2021} are interested in estimating the natural sky brightness at ground-based observing sites (including also zodiacal light, airglow and radiative transfer through the atmosphere); they do not discuss the LISRF {\em per se}.
Actually, an estimate on the LISRF in the optical, from a vantage point outside the interplanetary dust cloud and including both direct and scattered radiation, would have been possible using data from the {\em Pioneer} 10 and 11 probes \citep{WeinbergJGR1974}. However, the data had been used to study diffuse MW emission \citep{GordonApJ1998} and cosmic, extragalactic, emission \citep{MatsuokaApJ2011}, but not the full LISRF.

\begin{figure*}
\centering
\includegraphics[width=\hsize]{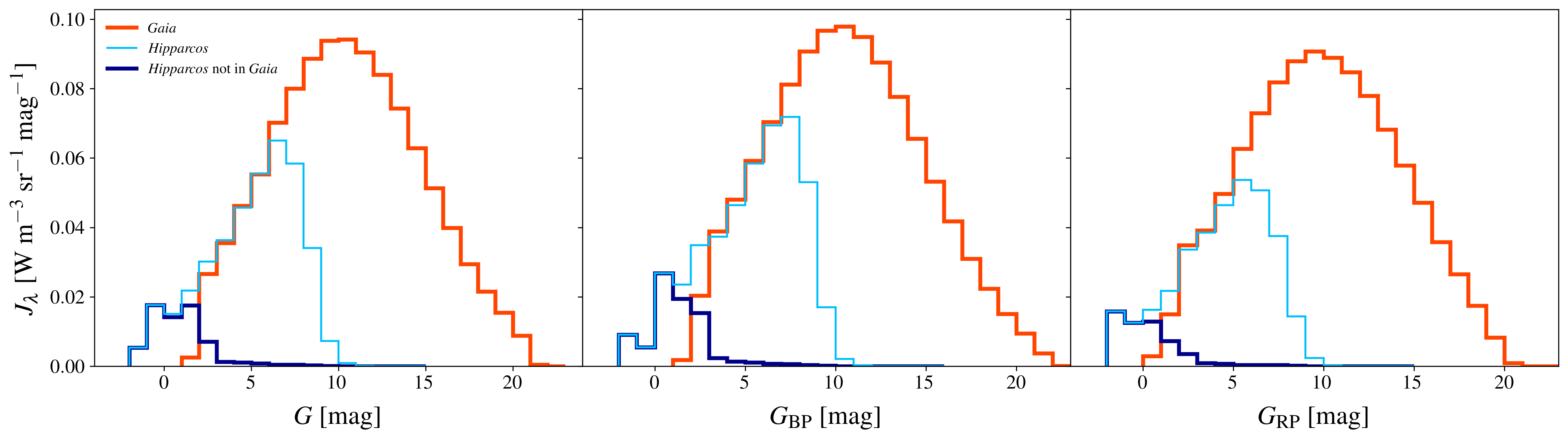}
\caption{
$J_\lambda$ vs {\em Gaia} magnitudes, for stars in the {\em Gaia} and {\em Hipparcos} catalogs,
and for the {\em Hipparcos} stars with no {\em Gaia} counterpart.
}
\label{fig:JvsG}
\end{figure*}

The purpose of this work is to derive a new estimate of the LISRF starting from the {\em Gaia} Data Release 3 catalog \citep[DR3;][]{GaiaDR3summary}. First, in Sect.~\ref{sect:phot}, I compute the optical LISRF from direct starlight in the {\em Gaia} photometric bands, using fluxes from {\em Gaia} (and {\em Hipparcos} for the brightest stars). In Sect.~\ref{sect:compa}, I compare the results with MMP and other data, including, in particular, new optical estimates from the {\em Pioneer} probes (the full derivation of which is presented in Appendix~\ref{app:newe}). In Sect~\ref{sect:spectrum}, I extrapolate the LISRF from the optical to the UV and NIR, using
{\em Gaia} DR3-based parameters for stellar atmospheres; and I validate the results against full sky data (and against selected catalog crossmatches in Appendix~\ref{app:xm}). The spectrum for diffuse radiation is derived in Sect.~\ref{sect:diffuse} using dust emission as a template. The full LISRF, including both direct and diffuse starlight, is presented and discussed in Sect.~\ref{sect:total}. 
As an application, in Sect.~\ref{sect:dustem}, I evaluate the impact that the new LISRF will have in the verification of MW dust models. Results are summarised in Sect.~\ref{sect:summary}.

\section{The LISRF in the {\em Gaia} photometric bands}
\label{sect:phot}

Expressing the LISRF in terms of average surface brightness, it is
\begin{equation}
J_\lambda = \frac{1}{4\pi}\int_{4\pi} I_\lambda \; d\Omega,
\label{eq:jnu}
\end{equation}
where $I_\lambda$ is the surface brightness from a specific direction in the celestial 
sphere\footnote{
Alternatively, the LISRF can be written in terms of energy density,
$u_\lambda={4\pi J_\lambda}/{c}$ \citep{DraineBook2011}.}.
In order to show the contribution to $J_\lambda$ of different parts of the sky, 
whole sky maps will be presented here, adopting the Hierarchical Equal Area isoLatitude Pixelization (HEALPix\footnote{http://healpix.sourceforge.net}; \citealt{GorskiApJ2005}). 
Since all 
HEALPix pixels have the same area $\Delta\Omega$, Eq.~\ref{eq:jnu} reduces to 
\begin{equation}
J_\lambda = \frac{1}{N_\mathrm{pix}} \sum_{i=0}^{N_\mathrm{pix}} I^i_\lambda,
\end{equation}
where $N_\mathrm{pix}$ is the total number of pixels in the map and $I^i_\lambda$
the surface brightness from each pixel. If one consider only direct starlight,
$I^i_\lambda = F^i_\lambda/\Delta\Omega$,
with $F^i_\lambda$ the total flux from all the stars within the pixel.
Eq.~\ref{eq:jnu} becomes
\begin{equation}
J_\lambda = \frac{1}{4\pi} \sum_{i=0}^{N_\mathrm{pix}} F^i_\lambda,
\end{equation}
which is equivalent to
\begin{equation}
J_\lambda = \frac{1}{4\pi} \sum_{k=0}^{N_\mathrm{stars}} f^k_\lambda,
\end{equation}
showing that $J_\lambda$ can be derived from a simple summation of the flux $f^k_\lambda$ of each star.
In this work, HEALPix maps are used of order 6 ($N_\mathrm{pix}=49152$, $\Delta\Omega\approx 2.5\times10^{-4}$ sr $\approx 0.84$ deg$^2$). 

In this section, I describe the derivation of $J_\lambda$ in the three bands of the {\em Gaia} Photometric Instrument: $G$, $G_\mathrm{BP}$ and $G_\mathrm{RP}$ ($\lambda_\mathrm{pivot}=$ 
0.62, 0.51, 0.78 $\mu$m, respectively; \citealt{GaiaDR3doc}), including all stars in
{\em Gaia} DR3 and the bright stars in the catalog of the
{\em Hipparcos} mission \citep{ESASP1997}, for which no {\em Gaia} photometry is available.

Here and in the rest of the paper, $I_\lambda$ and $J_\lambda$ will be expressed in units of W~m$^{-3}$ sr$^{-1}$ (equivalent to $10^{-3}$ erg cm$^{-2}$ s$^{-1}$ $\mu$m$^{-1}$ sr$^{-1}$). In these units, $I_\lambda$ and $J_\lambda$ in the optical range have values of order unity.

\subsection{Gaia}

The  DR3 contains photometric information for about 1.8 billion objects in $G$, 
and for about 1.5 billion in  $G_\mathrm{BP}$ and $G_\mathrm{RP}$ \citep{GaiaDR3summary}.
In order to construct the map of $F^i_\lambda$, I extracted from the
main source catalog\footnote{Unless otherwise specified, all {\em Gaia} 
products used in this work, as well as the {\em Hipparcos} catalog, were downloaded from 
the {\em Gaia} ESA Archive: https://gea.esac.esa.int/archive} 
({\tt gaia\_source}) the {\tt source\_id} field, which encodes the position in the HEALPix representation; and the {\tt phot\_g\_mean\_flux}, {\tt phot\_bp\_mean\_flux}, {\tt phot\_rp\_mean\_flux} fields, containing the stellar flux in the three bands. The native fluxes were converted 
to units of energy per unit time, surface and wavelength using the conversion factors provided 
in Sect. 5.4.1 of the DR3 documentation \citep{GaiaDR3doc}. 

The fluxes of bright stars were corrected for saturation effects, using the magnitude-dependent
formulas in \citet{RielloA&A2021}. These
corrections account for minimal variations in the final $J_\lambda$ values:
only 0.2\% in $G$, and about 1\% in $G_\mathrm{BP}$ and $G_\mathrm{RP}$. For simplicity,
I did not consider the $\sim$5 million stars with missing $G$ photometry in the main
photometric catalog, for which alternative estimates are given in a supplementary catalog
\citep{RielloA&A2021}: they would only account for variations in $J_\lambda$
of about 0.01\% in $G$, and about 0.1\% in $G_\mathrm{BP}$ and $G_\mathrm{RP}$.
Instead, I applied corrections for the $\sim$0.5 billion stars without 
$G_\mathrm{BP}$ and/or $G_\mathrm{RP}$ fluxes, following a procedure similar 
to what done by \citet{MasanaMNRAS2021}: I first derived the average $G_\mathrm{BP}$/$G$ and 
$G_\mathrm{RP}$/$G$ ratios within each pixel - using stars with the complete photometric dataset; 
then, I assigned a flux according to the $G$ value and the corresponding flux ratio.
Nevertheless, these objects contribute to only 0.7 and 1.2\% of  $J_\lambda$
in $G_\mathrm{BP}$ and $G_\mathrm{RP}$.

In Fig.~\ref{fig:JvsG}, I show the contribution of stars of different magnitude
to $J_\lambda$, for the three {\em Gaia} photometric bands. 
The maximum contribution to $J_\lambda$, in all bands, comes from stars
with magnitudes $\approx10$. For magnitudes $\la2$,the distributions shows a
break (most evident for $G$) due to the lack of very bright stars in the {\em Gaia} 
catalog: I will try to recover these using the {\em Hipparcos}
catalog (Sect.~\ref{sect:hipparcos}). On the faint-end side, the distributions drops 
smoothly: in $G$, it goes down to the DR3 completeness limit at $\approx21$ 
magnitudes \citep{GaiaEDR3summary}.
\citet{MasanaMNRAS2021} estimate the ratio between the total flux of fainter stars with 
$G>21$ and brighter stars with $10.5<G<20$, as a function of the position in the Galaxy:
judging from their Fig.~4, the contribution of fainter stars is, {\em at most}, 5\%
for lines of sight in the inner disk (for Galactic longitudes $|b|\le90^\circ$ and 
latitudes $|l|\le10^\circ$) and 0.05\% elsewhere. Using these value and a full sky
map for objects with $10.5<G<20$, I find that faint stars can contribute a
maximum of 1\% to $J_\lambda$ in the $G$ band;
this value is similar to the contribution of the last populated bin in Fig.~\ref{fig:JvsG}.
In principle, this value should be added to the calculation to derive an upper value for $J_\lambda$ in the $G$ band.
However, given the smallness of this contribution and its 
gross estimate, I prefer to avoid a cumbersome notation
and consider it instead as an uncertainty in the estimate of $J_\lambda$, with the caveat that the value presented here is biased low with respect to the full $J_\lambda$ including stars with $G>21$, but still compatible within  1\%.
Since the $J_\lambda$ distributions for $G_\mathrm{BP}$ and $G_\mathrm{RP}$
are very similar to that for $G$, I use the same 
uncertainty also for those bands.

A small fraction of the LISRF is be due to extragalactic objects. 
Estimates based on galaxy counts predict $J_\lambda \approx $ 
0.01 W~m$^{-3}$ sr$^{-1}$ over the bands considered here \citep{DriverApJ2016}.
If the potential contribution of distant galaxies to DR3 is selected, using the fields 
{\tt in\_galaxy\_candidates}, {\tt in\_qso\_candidates} and {\tt in\_andromeda\_survey}, 
I find similar values:
$J_\lambda =$ 0.008, 0.01 and 0.007 W~m$^{-3}$ sr$^{-1}$, for $G$, $G_\mathrm{BP}$ and $G_\mathrm{RP}$, respectively. However, there
is no guarantee on the completeness and purity of the DR3 extragalactic selections:
for example, the {\em Gaia} Andromeda Photometric Survey \citep{EvansA&A2023}
selects all objects - including Galactic stars - within 5.5$^\circ$ of M31; also,
I found an anomalous concentration of objects near the Galactic center - i.e.\ Galactic 
stars misclassified as extragalactic objects. Therefore,
the values above are likely upper limits to the true extragalactic contribution from
DR3. 
In this work, I retain the {\em Gaia} 'extragalactic' 
contribution in our estimates: it might contain a dominant fraction of 
stellar radiation and extragalactic fluxes, however underestimated, should be added 
to the local radiation field. 
On the other hand, adding the estimates from independent
galaxy counts will result in an upper limit to the total
$J_\lambda$, since a fraction of extragalactic objects
are indeed present in the {\em Gaia} catalog. Thus, in 
analogy to what done for faint stars, I will 
consider the value from galaxy counts as 
a further uncertainty on the total $J_\lambda$.

\begin{figure*}
\centering
\includegraphics[width=\hsize]{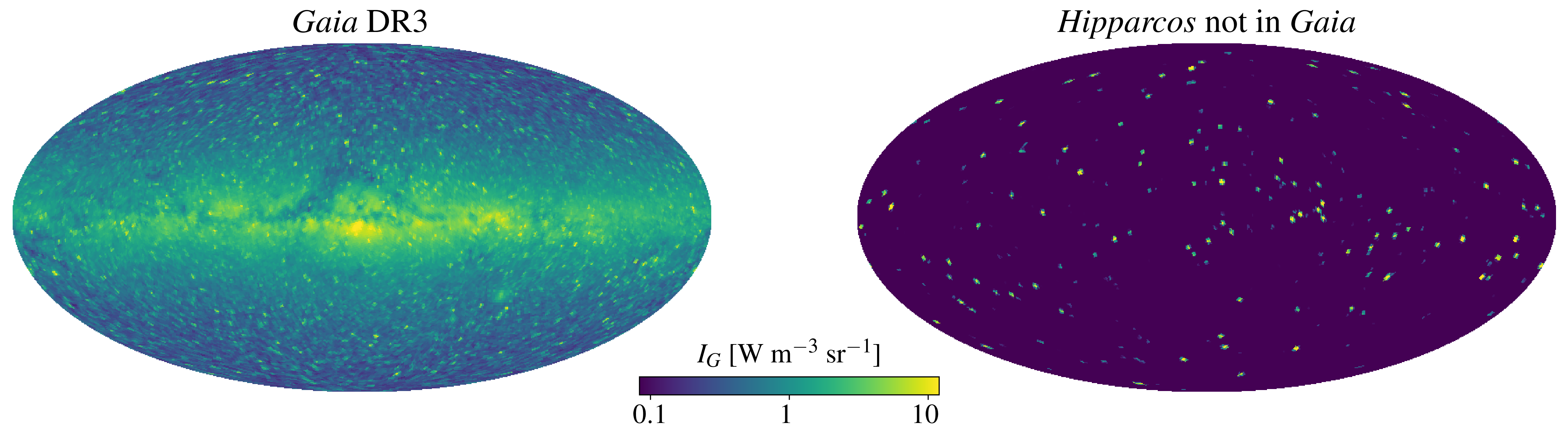}
\caption{
The $G$-band surface brightness $I_G$ from the {\em Gaia} DR3 catalog (left) and from the {\em Hipparcos} stars not in DR3  (right). 
To ease the visualization of the {\em Hipparcos} contribution, both maps are been smoothed with a Gaussian of FWHM equal to the pixel size.
All maps in this work are shown in Galactic coordinates and Mollweide projection
(the extent of each map being delimited by Galactic latitude $b=-90^\circ$ 
at bottom and $b=90^\circ$ at top, and by Galactic longitude $l=-180^\circ$ 
at right and $l=180^\circ$ at left, respectively).
}
\label{fig:I_G}
\end{figure*}

The surface brightness map computed from {\em Gaia} is shown in Fig.~\ref{fig:I_G}, for the {\em G} band; its average, the LISRF $J_\lambda$, is presented in Table~\ref{tab:J_l}, for the three {\em Gaia} bands.
The uncertainty results from the sum, in quadrature, of the uncertainties on the contribution of
very faint stars and extragalactic objects, discussed above; and of a further 1\%
uncertainty in the photometric absolute calibration \citep{GaiaDR3doc}. Instead, the total contribution of the photometric errors listed in the catalog is irrelevant, being an 
order of magnitude smaller than the other components I considered here.

\begin{table*}
\caption{Contribution of direct starlight to the LISRF in the {\em Gaia} bands, plus additional estimates for other samples.}
\label{tab:J_l}
\centering
\begin{tabular}{lccc}
                &\multicolumn{3}{c}{$J_\lambda$ [W~m$^{-3}$ sr$^{-1}$]}\\\\
                &$G$        &$G_\mathrm{BP}$       &$G_\mathrm{RP}$\\ \hline
\\
&\multicolumn{3}{c}{LISRF}\\\\
{\em Gaia} DR3&   1.07$\pm$0.02  &1.12$\pm$0.02    & 1.06$\pm$0.02   \\
{{\em Hipparcos} not in {\em Gaia}}
                &  0.07$\pm$0.01  & 0.08$\pm$0.01    & 0.06$\pm$0.01  \\
Total           &  1.14$\pm$0.02  & 1.20$\pm$0.02    & 1.12$\pm$0.02   \\
\hline
\\
&\multicolumn{3}{c}{other samples}\\\\
{\em Hipparcos} all&   0.39  & 0.46    & 0.34   \\
{\em Hipparcos} \& {\em Gaia} cross-match ({\em Hipparcos} photometry) &   0.266  & 0.301    & 0.238   \\
{\em Gaia} \& {\em Hipparcos} cross-match ({\em Gaia} photometry) &   0.260  & 0.297    & 0.239   \\
{\em Gaia} AP&   0.63  & 0.69    & 0.58   \\
\hline
\end{tabular}
\end{table*}

\subsection{Hipparcos}
\label{sect:hipparcos}

Very bright stars are not included in the {\em Gaia} catalog, because of saturation. 
I recover the contribution to  $J_\lambda$ of those objects using the {\em Hipparcos} 
catalog. 

First the photometry in the {\em Gaia} bands is derived for the 
$\sim$118000 {\em Hipparcos} stars: I used the catalog {\em H}$_\mathrm{P}$ photometry and {\em V-I} 
colors and the \citet{RielloA&A2021} transformations, after verifying that
{\em V-I} values are in the range of applicability of the formulas (for about 1200 stars with
missing  {\em V-I}, I used the mean value in the corresponding parts of the sky).

The $J_\lambda$ values for the whole {\em Hipparcos} catalog are about one third 
of the {\em Gaia} values (Table~\ref{tab:J_l}). A substantial part of these $J_\lambda$ must 
be due to stars already included in the previous estimate. According to the DR3 cross-match 
between the two catalogs, available from the Archive, $\sim$85\% of the {\em Hipparcos} stars 
has {\em best-neighbour} {\em Gaia} counterparts. The values of $J_\lambda$ for this subsample
are $\sim 65\%$ of the full {\em Hipparcos} estimates; those derived from the {\em Hipparcos}
photometry are within $\la 2\%$ of the corresponding values obtained directly from {\em Gaia} 
photometry, providing an indication on the uncertainties on the photometric transformations used 
here (Table~\ref{tab:J_l}; see also Appendix.~\ref{sect:gvsh}). 

In principle, one could isolate the contribution of stars missing from the {\em Gaia} catalog
by selecting the {\em Hipparcos} stars not included in the cross-match. This is the approach of
\citet{MasanaMNRAS2021}: using DR2 they conclude that these stars account for $\sim$20\% of 
the total stellar radiation in {\em G}.
However, \citet{MarreseA&A2019} argued that only a small fraction of {\em Hipparcos} 
stars should be really out of the {\em Gaia} catalog and that the majority of the missing counterparts 
in the cross-match is due to problems with the astrometric solutions. 
For an independent check on the overlap between {\em Hipparcos} and
{\em Gaia}, I also plot in Fig.~\ref{fig:JvsG} $J_\lambda$ for {\em Hipparcos}. 
If the bright and faint end sides are excluded and only 
the range $3<G<6$ is considered, the $J_\lambda$ distributions for {\em Gaia}  and {\em Hipparcos} 
are essentially the same, meaning that the same stars are present in both.
While the drop for, e.g. $G\ga7$, is expected due to the shallower completeness 
limit of {\em Hipparcos}, $J_\lambda$ for, e.g. $G\la2$, is almost entirely due to 
{\em Hipparcos} stars missing from {\em Gaia}. 

Beside the official best-neighbour crossmatch, \citet{MarreseA&A2019}  provide 
the results of a nearest-neighbour cone search of {\em Hipparcos} stars within 1" of
a {\em Gaia} star. About 1500 {\em Hipparcos} stars are missing from this 
alternative list (compared to the $\sim$19000 missing from the DR3 cross-match).
Here I assume that these stars are those truly missing from DR3 and for which photometry is available {\em only} 
from the {\em Hipparcos} catalog: indeed their $J_\lambda$ vs magnitude distribution 
is almost complementary to that of {\em Gaia} (Fig.~\ref{fig:JvsG}).

The surface brightness map resulting from the independent {\em Hipparcos} contribution 
is shown in Fig.~\ref{fig:I_G}, for the {\em G} band; the $J_\lambda$ values for all bands 
are presented in Table~\ref{tab:J_l}. The uncertainty on $J_\lambda$
was estimated by applying simple magnitude cuts to the {\em Hipparcos} catalog in 
correspondence to the bright-end drop of the {\em Gaia distribution} shown in 
Fig.~\ref{fig:JvsG}: it is about 15\% of the {\em Hipparcos} final contribution,
reflecting the uncertainty in the choice of the contributing stars (larger than the 
uncertainty in the photometric transformations discussed above). As evident from the 
very sparsely populated map in Fig.~\ref{fig:I_G},  
the {\em Hipparcos}-only contribution is dominated by a small number of 
nearby bright stars: for example, there are only 68 objects with $G<2$, constituting 
almost 90\% of the value of $J_\lambda$ for the {\em G} band listed in Table~\ref{tab:J_l}.

\vspace{1em}
\noindent The final LISRF from direct starlight, resulting from the combination of 
the {\em Gaia} and {\em Hipparcos} catalogs, is also given in Table~\ref{tab:J_l}.

\begin{figure*}
\centering
\includegraphics[width=\hsize]{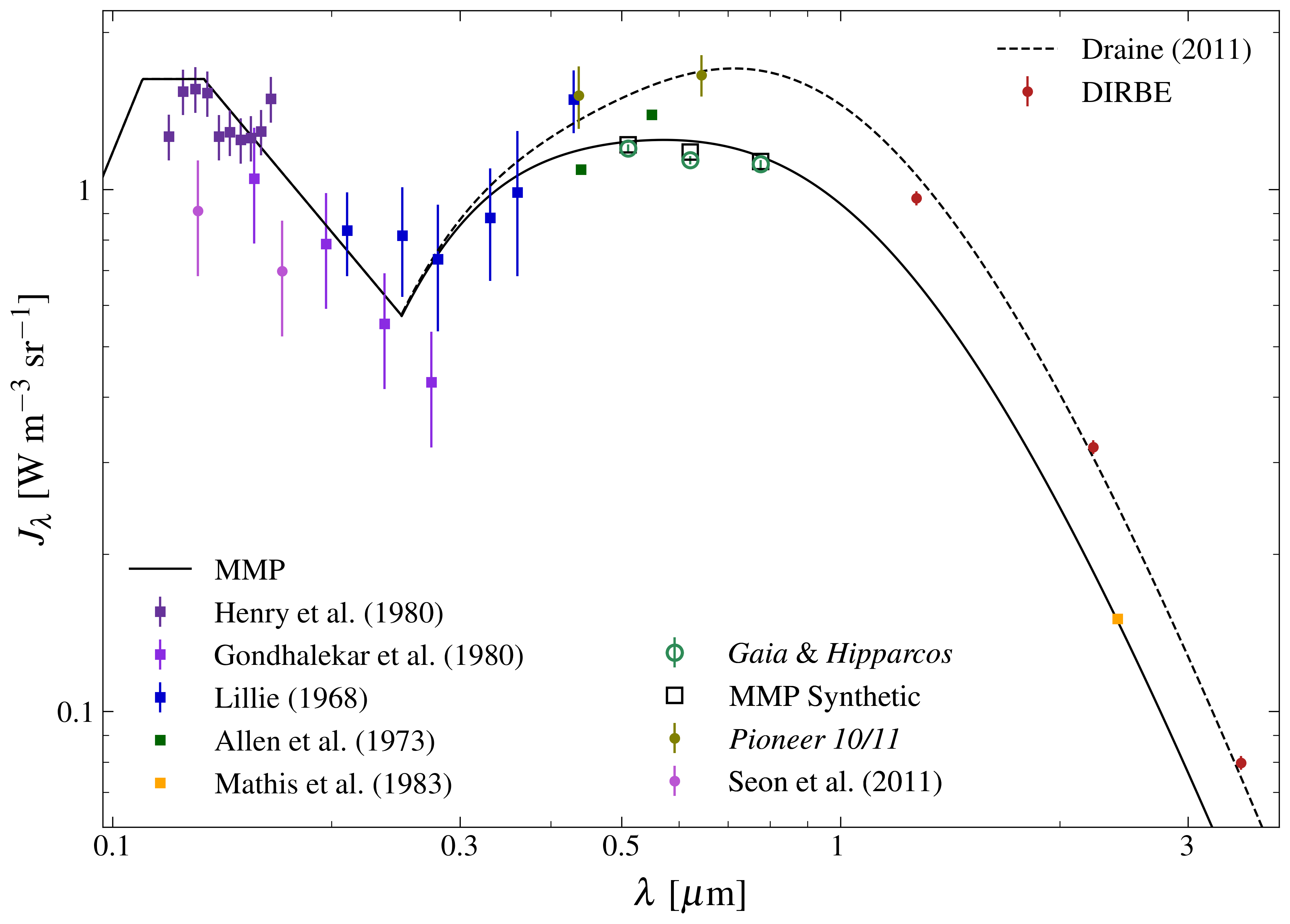}
\caption{
Comparison between the {\em Gaia/Hipparcos} estimates and MMP. I also show: the observational constraints of MMP (listed in the bottom-left legend); synthetic photometry of the MMP model over the {\em Gaia} bands; the \citet{DraineBook2011} update based on DIRBE full sky data; our estimates from {\em Pioneer 10/11} data; the more recent UV estimates from \citet{SeonApJS2011}. See main text and Appendix~\ref{app:newe} for details.
}
\label{fig:LISRF}
\end{figure*}

\section{Comparison with MMP and full-sky surveys}
\label{sect:compa}

The $J_\lambda$ values estimated in the previous section are compared with the MMP LISRF model in Fig.~\ref{fig:LISRF}.

In the optical-NIR range, the MMP model consists of the sum of three 'dilute' (i.e.\ 
scaled-down) blackbodies. 
In its original derivation \citep{MezgerA&A1982} the blackbodies were just two: the dilution factors were chosen to broadly reproduce near-UV estimates from satellite and B-band estimates 
from ground-based observations (from \citealt{LillieThesis1968}, 
as reported by \citealt{WittApJ1973}), and B and V-band estimates from 
stellar counts (as tabulated in \citealt{AllenBook1973}). \citet{MathisA&A1983} made a few modifications
and introduced a third dilute blackbody to match estimates of the LISRF in the NIR,
extrapolated from rocket-borne observations in the Galactic plane at 2.4~$\mu$m \citep{HayakawaPASJ1978}. Nevertheless, the values of $J_\lambda$
in the optical range changed little with respect to \citet{MezgerA&A1982}.
The MMP model is shown in 
Fig.~\ref{fig:LISRF}, together with the observational data from which it has been derived.
The model is very close to our estimates: when I use it to derive 
synthetic photometry in the three {\em Gaia} bands, I find a difference within $\la 4\%$,
much less than the 15\% uncertainty claimed by the authors. Thus, at least in the optical 
range,  MMP is a good representation of the direct starlight component of the LISRF.

The NIR component introduced by \citet{MathisA&A1983} proved insufficient to explain the LISRF 
resulting from full-sky DIRBE observations \citep{ArendtApJ1998}. In order to
match the newer NIR data, \citet{DraineBook2011} used the two dilute blackbodies of \citet{MezgerA&A1982} and 
increased the dilution factor of the third blackbody of \citet{MathisA&A1983}.
To my knowledge, the DIRBE values
are only shown graphically in Fig.~12.1 of \citet{DraineBook2011} and are not provided numerically
elsewhere: I estimated them directly from the maps (see Appendix~\ref{app:dirbe}
for the derivation). The \citet{DraineBook2011} LISRF and the DIRBE estimates are 
also shown in Fig.~\ref{fig:LISRF}. 
Beside roughly matching the DIRBE datapoints,
the new LISRF model is significantly brighter in the optical: when integrated over the {\em G} band,
$J_\lambda$ is $\sim40\%$ higher than the corresponding {\em Gaia/Hipparcos} value (the
synthetic {\em Gaia} photometry  for \citealt{DraineBook2011} is not shown in the figure).

Both MMP and the \citet{DraineBook2011} LISRF share the same spectrum for
$\lambda\la0.25 \mu$m. Below the hydrogen ionisation threshold, the spectrum is described
as a broken power law which \citet{MezgerA&A1982} scaled 
on a few $J_\lambda$ estimates: in the far-UV, using data for about 1/3 of the sky 
(but extrapolated to the whole celestial sphere) taken
with a spectrometer onboard the {\em Apollo 17} spacecraft \citep{HenryApJ1980}; for 
$0.15\la\lambda/\mu\mathrm{m}\la0.3$, from a full-sky survey from the TD-1 satellite 
\citep{GondhalekarA&A1980}; for $0.2\la\lambda/\mu\mathrm{m}\la0.5$, from observations of 
selected areas from the OAO-2 satellite \citep[plus additional optical ground-based observations;][]{LillieThesis1968}. The UV $J_\lambda$ values 
are shown in Fig.~\ref{fig:LISRF}, where I adopted as errors the calibration 
uncertainty quoted by \citet{HenryApJ1975} for the far-UV spectrometer, the
total uncertainty (dominated by calibration) quoted by \citet{GondhalekarA&A1980} 
for the TD-1 estimates, and the spread in the determinations of \citet{LillieThesis1968}
shown by \citet{WittApJ1973}. In a more recent work, \citet{SeonApJS2011} derived $J_\lambda$ 
using far-UV data for about $80\%$ of the sky, from the Far-Ultraviolet Imaging Spectrograph (FIMS) on the
STSAT-1 satellite: their values are lower than previous estimates (see Fig.~\ref{fig:LISRF},
where I plot the extremes of a rather featureless spectrum), but still marginally compatible within 
the large calibration error of the instrument \citep{EdelsteinApJL2006}. 
Some of the UV and NIR observational constraints will be also used from the next Section, 
where the $J_\lambda$ spectrum is extrapolated beyond the {\em Gaia} bands.

An advantage point for the measurement of the sky radiation field was that of the {\em Pioneer 10} and {\em 11} probes. The Imaging Photopolarimeters (IPPs) aboard these spacecrafts made several measurements, covering $\approx 90$\% of the sky, in two photometric bands in the optical range.
In their cruise to outer space, the two probes went beyond the interplanetary dust cloud, thus
avoiding one of the sources of uncertainties in the estimate of the LISRF from near-Earth positions: zodiacal light. The {\em Pioneer 10/11} data have been used to study the
diffuse galactic and extragalactic light \citep{TollerA&A1987, TollerProc1990, GordonApJ1998, MatsuokaApJ2011}.
The data has been distributed after removing the fluxes of bright stars ($V<6.5$). Thus,
in order to recover the full LISRF, this contribution has to be added back; this is done
in Appendix~\ref{app:pioneer}. The total {\em Pioneer} estimates are shown in Fig.~\ref{fig:LISRF};
they include both 
direct starlight and diffuse light scattered by dust in the ISM (diffuse starlight is
also included in the UV and NIR full-sky estimates discussed above).
The {\em Pioneer}-based $J_\lambda$ values are higher (by $\sim$40\% in $G$)
than the starlight-only estimates from {\em Gaia/Hipparcos} (or MMP), 
highlighting the necessity of including diffuse radiation in the LISRF evaluation; 
this will be done in Sect.~\ref{sect:diffuse}.

\section{The LISRF spectrum for direct starlight}
\label{sect:spectrum}

In this Section, I use models of the spectra of {\em Gaia}
and {\em Hipparcos} stars to extrapolate $J_\lambda$ beyond the {\em Gaia} bands.
For each star, the observed (i.e.\ dust-attenuated) spectrum $f_\lambda$
can in principle be derived from
the observed flux in the {\em G} band
if we have a good model of the 
intrinsic, un-attenuated,
spectrum $\tilde{f_\lambda}$ and the reddening is known.
Assuming for the moment that observations are done at a specific wavelength (e.g.
the central wavelength of the $G$ band)
and not integrated over a filter bandwidth, 
we can write
\begin{equation}
f_\lambda
=f_\mathrm{G}
\,
\left(
\frac{\tilde{f}_\lambda}{\tilde{f}_G}
\right)
\times
10^{-0.4\times (A_\lambda-A_G)},
\label{eq:spectrum}
\end{equation}
where $f_G$ is the flux at 
the central wavelength of band $G$
and 
$\tilde{f}_\mathrm{G}$ the flux that would be
observed in the absence of dust extinction.
The dust extinction $A_\lambda$ can 
be written as the product of $A_0$, the 
extinction at a reference
wavelength, and $E_\lambda$, the 
extinction law normalised at 
the same wavelength;
analogously, $A_G$ is the 
dust extinction at 
the central wavelength of band $G$. 
In practice,
Eq.~\ref{eq:spectrum}
is the result of three passages:
first, the observed $f_\mathrm{G}$
is dereddened for extinction $A_G$
to obtain the intrinsic flux; second, the 
full intrinsic spectrum is recovered by 
scaling the flux model $\tilde{f}_\lambda$
to the intrinsic $G$ flux; finally, the full observable
spectrum is obtained by correcting
the unreddened spectrum for the 
reddening $A_\lambda$ at each $\lambda$.

Eq.~\ref{eq:spectrum} can be easily
generalized to the case of a broad 
filter by defining
\begin{equation}
f_G=
\int_G 
{f}_\lambda\,
S_\lambda\,  
d\lambda
\quad \mathrm{and} \quad
\tilde{f}_G=
\int_G 
\tilde{f}_\lambda\,
S_\lambda \,
d\lambda,
\end{equation}
where the integration is over
the filter band and $S_\lambda$ is the
filter response function; and
\begin{equation}
A_G=
-2.5\log_{10}
\int_G 
\frac{\tilde{f}_\lambda}{\tilde{f}_G}\,
S_\lambda \times\,
10^{-0.4 \times A_\lambda}\, d\lambda.
\label{eq:ag}
\end{equation}
For very narrow $S_\lambda$
(like, e.g., a delta function), $A_G$ 
tends to the extinction at the filter central wavelength. For a broad filter, such as the {\em Gaia G}
band, $A_G$ will instead be higher for intrinsically bluer stars and lower for redder ones. 

In the following, I use Eq.~\ref{eq:spectrum} to derive 
${f}_\lambda$ for each star in a catalog and sum their individual contribution
to obtain $I_\lambda$ maps and $J_\lambda$.
For  $\tilde{f}_\lambda$, I use the
library of stellar atmospheres of \citet{CastelliProc2003};
for $E_\lambda$, the average ($R_V=3.1$) MW
extinction law
of \citet{FitzpatrickPASP1999}. The spectrum is sampled (and averaged) over a grid of 99 wavelength bins logarithmically spaced between 0.1 and 4.3 $\mu$m.

\subsection{Gaia}
\label{sect:gaia_spectrum}

The DR3 provides Astrophysical Parameters (APs) for about 470 million stars, derived by fitting {\em BP} and {\em RP} low resolution spectra with an extended library of various synthetic 
stellar atmosphere models \citep[for details, see][]{AndraeA&A2023,CreeveyA&A2023}. I use here a subset of AP quantities, obtained from the 
best-fitting spectral library and available in the main {\tt gaia\_source} catalog.

I derived the unattenuated stellar spectrum, $\tilde{f}_\lambda$, from the effective temperature $T_\mathrm{eff}$ ({\tt teff\_gspphot}),  the surface gravity $\log g$ ({\tt logg\_gspphot}), and 
the metallicity [M/H] ({\tt mh\_gspphot}).
About 90\% of the fitted parameters are within the grid boundaries of the 
\citet{CastelliProc2003} library adopted here (the original libraries used for the APs derivation are not publicly available); for the rest, I assume the values at the nearest boundary. In practice, I
pre-computed spectra over a table spanning 100 logarithmically-spaced temperature values within 
2400~K$<T_\mathrm{eff}<$42000~K; 25 $\log g$ values within $-0.5<\log g<5.5$;
45 [M/H] values within $-4.5<\mathrm{[M/H]}<1$ (those are the boundaries of the AP estimates).
When deriving the spectrum for each star, I simply searched for the nearest entry in the table:
for the adopted grid, there is virtually no difference between this and a table interpolation. 

The dust-extinguished spectrum is derived according to Eq.~\ref{eq:spectrum}, using the
monochromatic extinction $A_0$ at 0.5477 $\mu$m ({\tt azero\_gspphot}), the $G$-band 
extinction $A_G$ ({\tt ag\_gspphot}), and the average \citet{FitzpatrickPASP1999} extinction law
(also used for AP estimates). While the results shown here use the $A_G$ from APs, there is little difference if the same quantity is derived from Eq.~\ref{eq:ag} computed over our spectrum grid
(plus an additional grid in $A_0$).

\begin{figure*}
\centering
\includegraphics[width=\hsize]{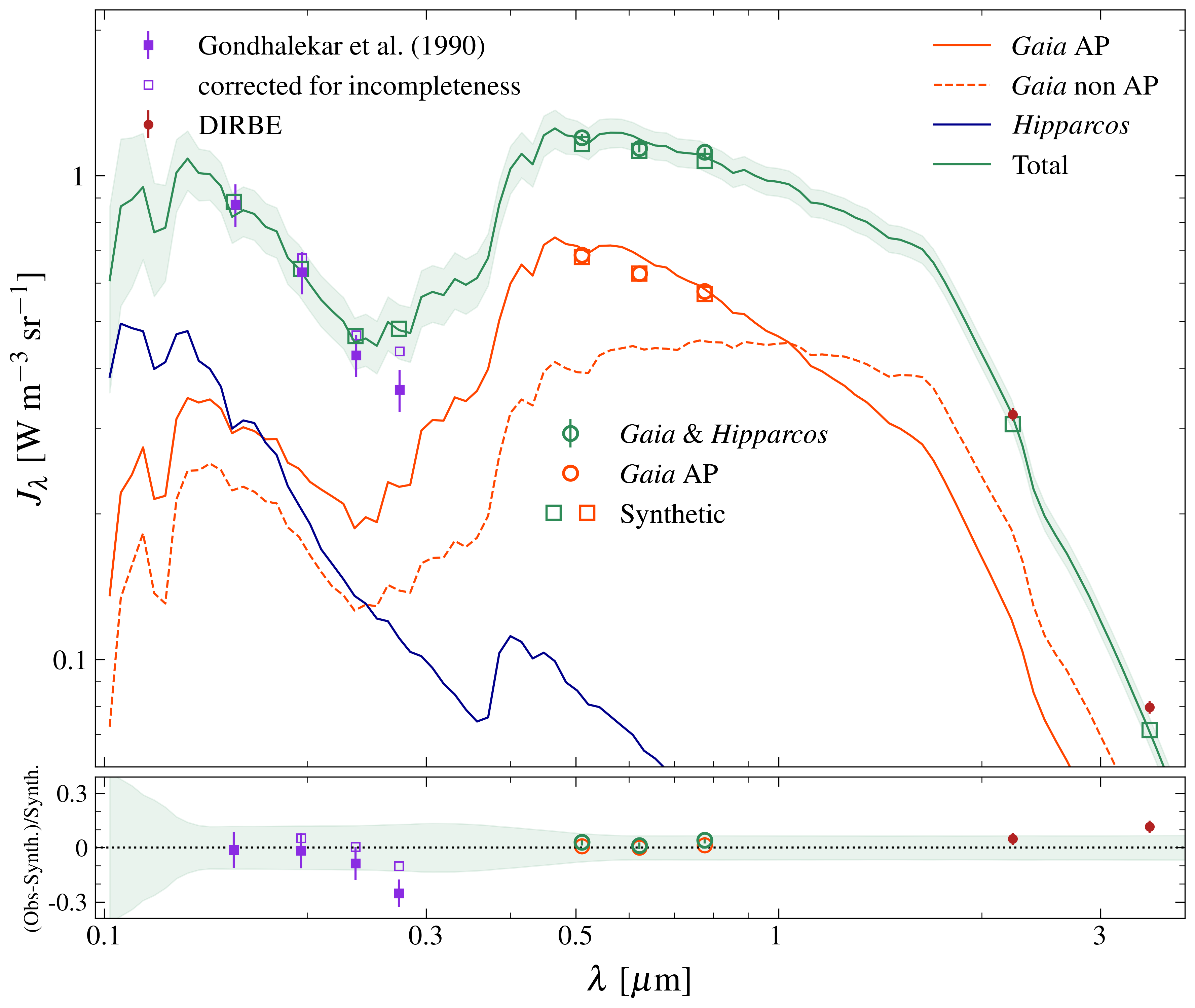}
\caption{
The diffuse starlight component of the LISRF spectrum. In the upper panel I show the total spectrum, the spectrum of {\em Gaia} AP stars, 
the recovered spectrum of {\em Gaia} non-AP stars, and the spectrum of the {\em Hipparcos} stars
not included in the {\em Gaia} catalog. For the total spectrum (and for the {\em Gaia} AP component)
I also plot the synthetic $J_\lambda$ values in the {\em Gaia} bands together with the estimates 
from the direct flux summation. I also show observational UV and NIR
estimates
and the corresponding synthetic photometry (for the full spectrum only).
The relative difference between observations and synthetic photometry is shown in the bottom panel.
The shaded area refers to the estimated uncertainty for the total spectrum (see text for details).
}
\label{fig:LISRFs}
\end{figure*}

In Fig.~\ref{fig:LISRFs} I show the final spectrum resulting from {\em Gaia} stars with available APs. I also include 
$J_\lambda$ derived from the direct summation of the fluxes of AP stars in the three
{\em Gaia} bands (values are presented in Table~\ref{tab:J_l}); and the synthetic photometry obtained by integrating the spectrum over the filter bandpasses. While the agreement in the $G$ band is almost by construction, the closeness between the $G_{BP}$ and $G_{RP}$  values and the synthetic
photometry was not granted, given the differences between the
AP stellar libraries and the one adopted here.

\begin{figure}
\centering
\includegraphics[width=\hsize,trim={0 0 26cm 0},clip]{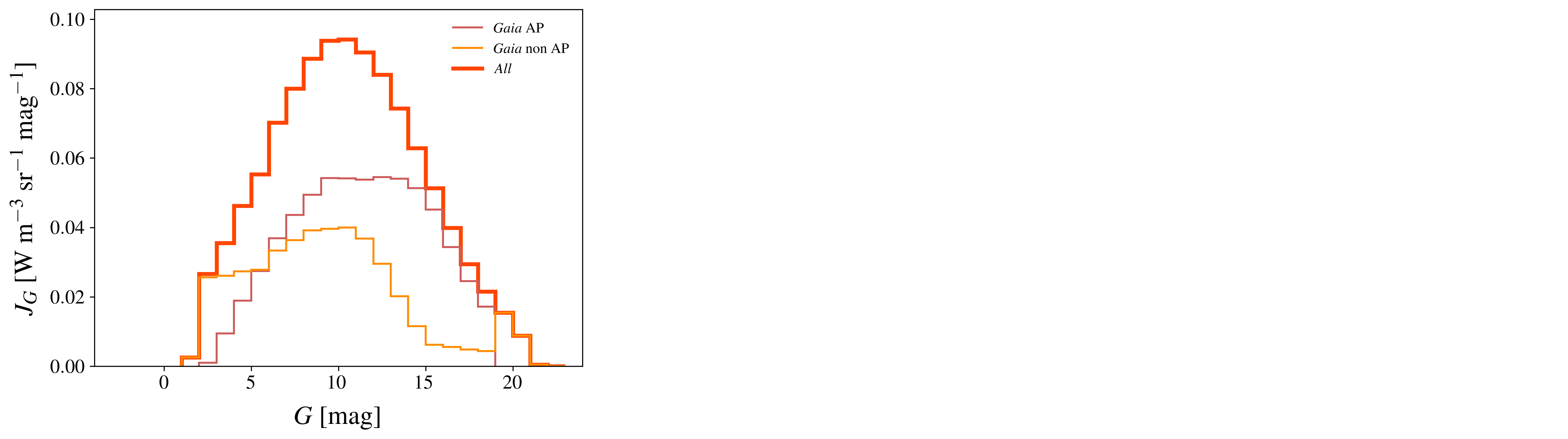}

\includegraphics[width=\hsize,trim={0 0 26cm 0},clip]{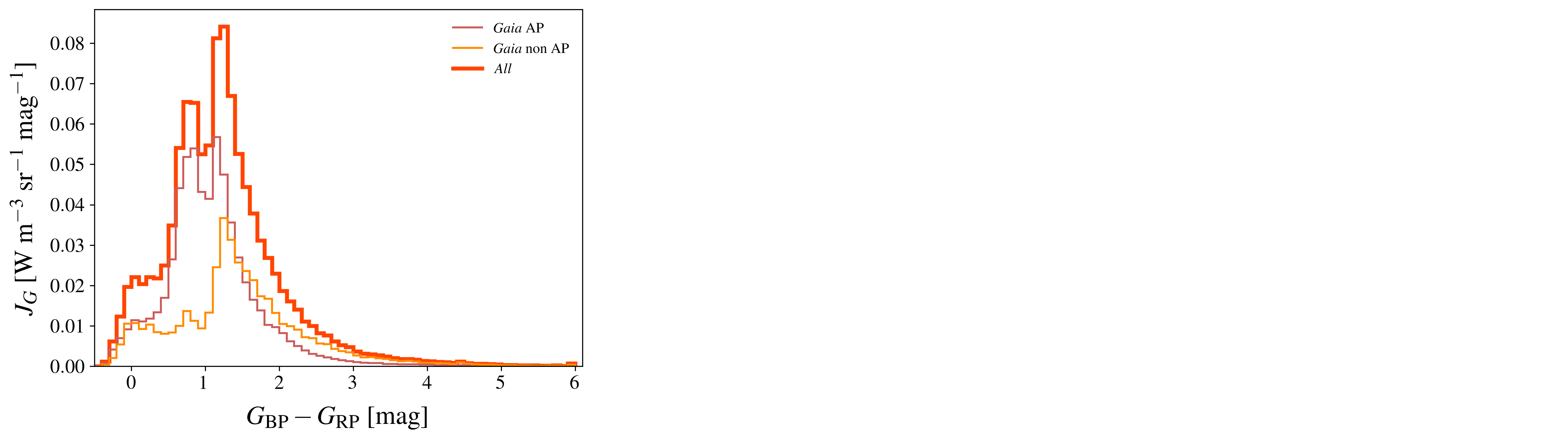}
\caption{
$J_G$ vs $G$ (top) and vs $G_\mathrm{BP}-G_\mathrm{RP}$ (bottom), for {\em Gaia} stars with and without Astrophysical Parameters.
}
\label{fig:JvsBR}
\end{figure}

The intensity and color ratios of the photometric $J_\lambda$ for the AP selection are clearly different from those of the full stellar derivation of the previous Section (also shown in  Fig.~\ref{fig:LISRFs}) or from the values for Gaia stars only (from Table~\ref{tab:J_l}).
This is due to the complex selection applied to the full {\em Gaia} sample before producing the AP catalog. First, the fit was not attempted for stars with $G>19$, amounting to 2/3 of {\em Gaia} entries; the fits  on the remaining objects were further filtered before publishing 
the catalog, leaving out about 100 thousand stars, preferentially of brighter $G$ and lower 
signal-to-noise in the parallax determination
\citep{AndraeA&A2023}.
In Fig.~\ref{fig:JvsBR} we show the separate contribution of AP and non-AP stars to $J_G$, as a function of the $G$ magnitude and of the $G_\mathrm{BP}-G_\mathrm{RP}$ color.
The distribution with magnitude (top panel) shows that AP stars contributing most to $J_G$ are on average
fainter than non-AP stars, lacking in particular the very-bright stars: this is the reason of the reduced $J_\lambda$ values. Instead,
the omission of $G>19$ stars has, as expected, little 
impact on the total values.
The distribution with colors of all {\em Gaia} stars (bottom panel) shows two main peaks: stars with the largest contribution to $J_G$ have $G_\mathrm{BP}-G_\mathrm{RP} \approx 1.2$, followed by those with $G_\mathrm{BP}-G_\mathrm{RP} \approx 0.8$.
In the AP selection, the redder peak is preferentially suppressed with respect to the bluer one, probably as a result of the selection against poorer parallax measurements, leaving out redder, more distant stars.
The net effect is that the photometric $J_\lambda$
values for AP stars (and the corresponding spectrum) are bluer than for the full catalog.


The final step is to recover the
spectrum of {\em Gaia} stars without APs.
I adopted for this a simple procedure: 
i) using the binning in $G_\mathrm{BP}-G_\mathrm{RP}$ shown in Fig.~\ref{fig:JvsBR}, I derived the average spectrum from all AP stars in each color bin  and each
pixel, and normalize it to the corresponding $J_G$ value;
in case there where no AP stars in a particular bin for a given pixel, I used the 
whole-sky average for that color bin;
ii) I assume that the same spectral shape is valid on average for the contribution of
non-AP stars in the same color bin and pixel;
iii) I scale the average spectrum by the $J_G$ contribution of non-AP stars.
The resulting spectrum is shown in Fig.~\ref{fig:LISRFs}: as expected, it is redder than $J_\lambda$ from AP stars. 


\subsection{Hipparcos}
\label{sect:hipparcos_spectrum}

For {\em Hipparcos} stars I followed \citet{SeonApJS2011} and derived $T_\mathrm{eff}$
and $\log g$ from spectral types, using the tables of \citet{StraizysAp&SS1981};
in this case I assumed solar metallicity. The spectral types are taken from the
{\em Hipparcos} catalog and supplemented with data from the SIMBAD database\footnote{http://simbad.cds.unistra.fr/simbad/}, after clearing
peculiar entries. Whenever available  from SIMBAD, direct $T_\mathrm{eff}$ and $\log g$ estimates are preferred;
at the time I accessed the service (May 2023), they did not include {\em Gaia} AP
values and are thus independent.
The full spectral classification (either from spectral types or direct estimates)
is available for $\sim65$\% of the catalog. For an additional $\sim34$\%
the spectral type is not complete, missing (in the vast majority of cases)
the luminosity class of the star: in this case I draw it randomly from the 
sample with the full description. The rest of the objects (1\%) are assumed
to be stars of solar spectral type (G0V). The final result of this procedure is 
the association of each {\em Hipparcos} star to a table of $\sim$190 spectra, 
again derived from the \citet{CastelliProc2003} library over the adopted wavelength grid. 

The monochromatic extinction $A_0$ is derived by comparing observed stellar colors and the intrinsic spectra assigned to a star. In practice, I 
used the spectrum to compute synthetic colors as a function of extinction; then I select the $A_0$ value for which the synthetic color matches the observed one. For each star, I derive up to six $A_0$ values, using as colors 
$B-V$, $V-I$, $B_\mathrm{T}-V_\mathrm{T}$, $H_\mathrm{P}-I$, $V-J$, $I-K$ ($B$, $V$, $I$, $B_\mathrm{T}$, $V_\mathrm{T}$, $H_\mathrm{P}$ from the {\em Hipparcos} catalog, $J$ and $K$ from SIMBAD, the full set being available for 95\% of Hipparcos stars): I use
as $A_0$ the average of these estimates. For each spectrum and $A_0$, $A_G$ is derived using Eq.~\ref{eq:ag}. 
Finally, the spectrum $f_\lambda$
is computed for each star using Eq.~\ref{eq:spectrum} and the
$f_G$ value derived in Sect~\ref{sect:hipparcos}.

The $J_\lambda$ spectrum for the {\em Hipparcos} stars not included in the {\em Gaia} catalog is shown in Fig.~\ref{fig:LISRFs} (other spectra from different subsets of the catalog are presented in Appendix~\ref{app:xm}).

\begin{figure*}
\centering
\includegraphics[width=\hsize]{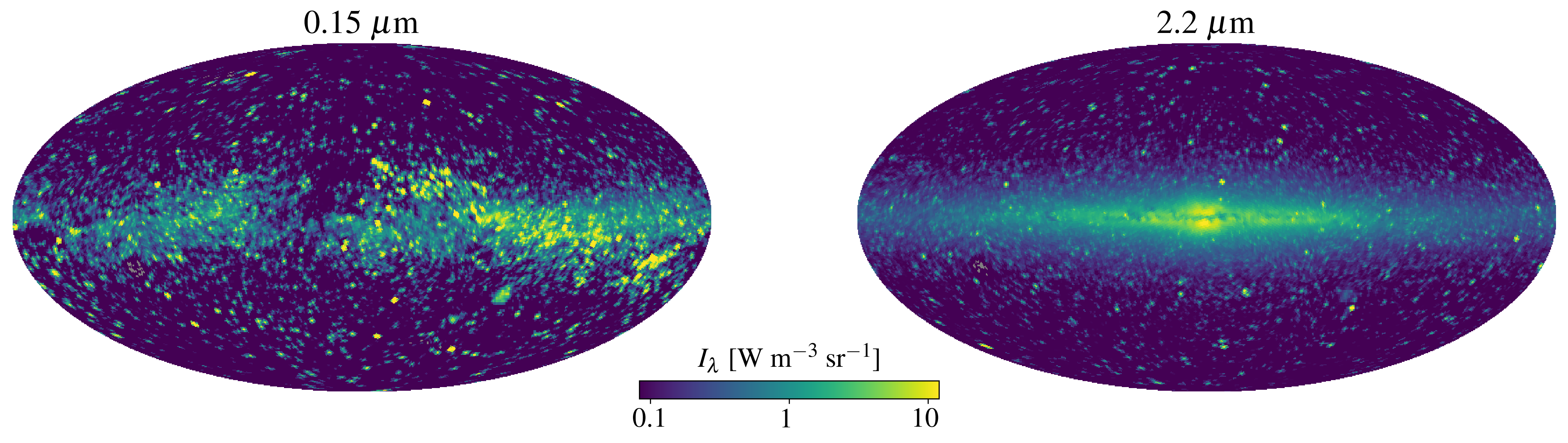}
\caption{
Predicted $I_\lambda$ from direct starlight at 0.15 $\mu$m and 2.2 $\mu$m. Maps have been smoothed with a Gaussian of FWHM=0.7$^\circ$ to match the resolution of DIRBE (and approximately that of the map in \citealt{SeonApJS2011}).
}
\label{fig:predicted}
\end{figure*}

\subsection{Results}

Several pitfalls can affect the extrapolation of stellar spectra discussed in the previous
sections: the APs are derived from optical data only and might be affected
by the extinction-temperature degeneracy \citep{AndraeA&A2023}; the spectral
libraries used for the AP fitting are not the same as the \citet{CastelliProc2003} library 
I use; the average MW extinction law might not be adequate along all sight-lines, in 
particular in the UV. 
The caveats on the extinction law and on the extinction-temperature degeneracy apply both to the {\em Gaia} and {\em Hipparcos} derivation; for the latter, there is also
the additional issue that the stellar parameters and extinctions come from heterogeneous sources and thus have not been derived simultaneously and self-consistently.
Despite all these (and possibly other) sources of uncertainties,
I find that the extrapolated spectra are still reliable, at least for $J_\lambda$ resulting
from the contribution of a large number of stars. The verification of the method is presented
in appendix~\ref{app:xm}: in short, I have selected various
cross-matches between independent photometric catalogs and the {\em Gaia/Hipparcos} ones
and compared $J_\lambda$ derived from UV, optical and NIR photometry with synthetic 
photometry in the same bands, from the extrapolated spectra. 
Here I assume 
as relative error for the total {\em Gaia} spectrum (AP stars plus non-AP extrapolation) the relative
error found for the samples analyzed in appendix~\ref{app:xm}; and the corresponding one for
{\em Hipparcos}. The final uncertainty is the sum in quadrature of both contributions.

The total $J_\lambda$ from direct starlight is shown in Fig.~\ref{fig:LISRFs}, together with its uncertainty.
Synthetic photometry in the {\em Gaia} bands compares well with the direct summation of {\em Gaia}
and {\em Hipparcos} fluxes, but not as well as for {\em Gaia} AP stars only: the difference in $G$ 
is still within the error quoted in Table~\ref{tab:J_l}, but about twice in the $G_\mathrm{BP}$  
and $G_\mathrm{RP}$ bands. The reason of the larger uncertainty is in the gross estimate of the  spectrum of {\em Gaia} stars without APs. Nevertheless, the difference is well within the conservative estimate of the global $J_\lambda$ uncertainty, which is about $7\%$ in the optical and NIR.

An estimate of $J_\lambda$ in the UV is obtained by \citet{GondhalekarProc1990} by 
summing the fluxes from stars observed by the TD-1 satellite. He used a
revised version of the original catalog by \citet{ThompsonBook1978},
with recalibrated fluxes resulting in a calibration uncertainty of 10\%
(about half of that quoted for the full -- direct + diffuse starlight -- estimate of \citealt{GondhalekarA&A1980}).
Synthetic photometry from the spectrum is in good agreement with the estimate from the 
TD-1 catalog, except at 0.274 $\mu$m. However, that flux in particular might be affected by 
incompleteness in the catalog. If I correct the \citet{GondhalekarProc1990} $J_\lambda$
for the incompleteness factors estimated in that work, all UV values agrees with the starlight
spectrum within their errors, and within the current estimated uncertainty, $\sim$13\% in the UV.
Apparently, {\em Gaia} \& {\em Hipparcos} stars account for all the UV starlight emission. 
In particular, the limited number of very bright and nearby stars present only in the {\em Hipparcos} catalog  contribute to almost half of $J_\lambda$ in this range.
The important contribution of nearby stars is also evident in the simulated $I_\lambda$ map at 
0.15 $\mu$m (i.e.\ the GALEX FUV band, one of the TD-1 bands, or the $L$-band in \citealt{SeonApJS2011}), shown in Fig.~\ref{fig:predicted}. Despite lacking the contribution of diffuse starlight, the map show similar features as those detected in far-UV surveys
\citep{HenryApJ1977,GondhalekarA&A1980,SeonApJS2011}: an 
asymmetry in Galactic longitude along the disk, between the $0^\circ<l<180^\circ$ 
and the brighter $-180^\circ<l<0^\circ$ ranges; the distribution of OB stars in the
Gould Belt, inclined with respect to the Galactic plane and particularly evident in the second longitude range (compare the map with the sky distributions of Gould Belt stars from \citealt{DeZeeuwAJ1999}, as shown by \citealt{HohleNewAR2008}). The same features are visible
also in optical $I_\lambda$ maps derived from the {\em Hipparcos} catalog (e.g. the $V<6.5$ direct starlight map in Fig.~\ref{fig:I_BIPP}), confirming the local origin of most of the LISRF in the UV.

\begin{figure}[b]
\centering
\includegraphics[width=\hsize]{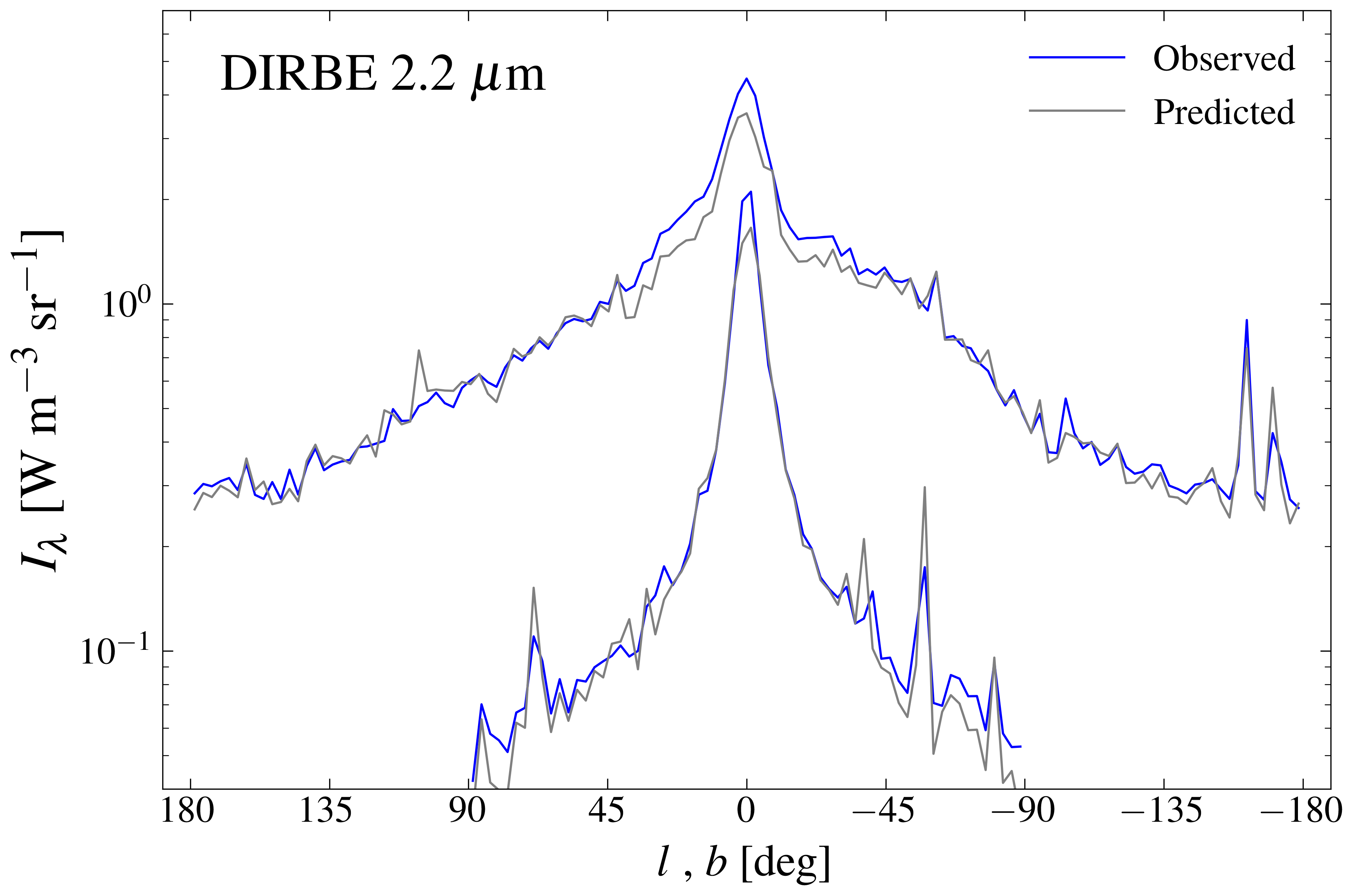}
\caption{
Profiles in the DIRBE 2.2 $\mu$m band, along Galactic longitude $l$ (averaged for $|b|<15^\circ$) and Galactic latitude $b$ (averaged over the whole $l$ range). 
}
\label{fig:lat_dirbe2}
\end{figure}

In the NIR, the DIRBE $J_\lambda$ values can be used as estimates of the LISRF from direct starlight, if observations are
excluded at 1.25 $\mu$m, where the contribution from dust-scattered radiation might be significant (see next section). 
In Fig.~\ref{fig:predicted} I show the predicted $I_\lambda$ map in the DIRBE 2.2 $\mu$m band. 
The $I_\lambda$ distribution on the plane in front of the Galactic bulge shows a depression (correspondent to the optical extinction lane) which is not as
evident in the map from 
observations, at least at the DIRBE resolution (Fig.~\ref{fig:I_D1}). The difference is also shown by the Galactic longitude (for $|l|<45^\circ$) and latitude  (at $b\approx0^\circ$) profiles of Fig.~\ref{fig:lat_dirbe2}; it  can be explained by the presence, in the central part of the disk, of
stars heavily extinguished in the optical and with emission peaking in the NIR, 
which are missing from the {\em Gaia} catalog (and thus from the current predictions; see also \citealt{HobbsExA2021}).
Despite this, the missed population does not seem to contribute significantly to $J_\lambda$: 
the difference between the current estimate and observation at 2.2~$\mu$m is of the same
order of the adopted NIR uncertainty (Fig.~\ref{fig:LISRFs}). The agreement also shows 
that the modeling of the spectrum of non-AP stars is not substantially incorrect.
At 3.5 $\mu$m, instead, the observed $J_\lambda$ is 13\% higher
than the prediction, almost twice the reference uncertainty. This could be attributed
in part to the missing population described above and the uncertainties in the modeling;
and in part to dust emission, which starts to be significant at this wavelength
\citep{ArendtApJ1998}, mainly because of the 3.2 $\mu$m feature 
of Polycyclic Aromatic Hydrocarbons \citep{DraineBook2011}.
\citet{SanoApJ2016} studied the various contributions to the emission detected by 
DIRBE and found that diffuse Galactic light at 3.5 $\mu$m 
accounts for about 5\% of total starlight, a correction that would make 
the $J_\lambda$ estimate marginally consistent with the adopted uncertainty. However, their result
was obtained for high Galactic latitude only, where the uncertainties due to the contribution
of zodiacal light in emission are high. Therefore, I no dot apply this correction and simply
take note of this difference.

Overall, the $J_\lambda$ spectrum for direct starlight shown in Fig.~\ref{fig:LISRFs}  
reproduces rather 
satisfactorily the (few) observational constraints.

\section{The LISRF spectrum for diffuse radiation}
\label{sect:diffuse}

Diffuse Galactic radiation scattered by dust has received more attention than the total 
LISRF, because it allows to study dust scattering properties and to access the lower layer 
of emission, the extragalactic background light. Isolating its contribution suffers from the same problems as the full LISRF study, that is terrestrial airglow and zodiacal light, plus the additional
burden of subtracting direct starlight (either via catalogs, removal of point sources, or selection of blank sky areas).
A way to overcome these issues is to 
study the fraction of radiation that spatially correlates with 
an independent tracer of dust, such as emission in the FIR \citep{BrandtApJ2012}. Correlations between diffuse radiation and
emission at $100\,\mu$m observed by the IRAS satellite,
$I_\nu(100\,\mu\mathrm{m})$, have been studied from the
UV to the NIR by several authors \citep{MurthyApJ2010,AraiApJ2015,KawaraPASJ2017,ChellewApJ2022}.

\begin{figure}
\centering
\includegraphics[width=\hsize]{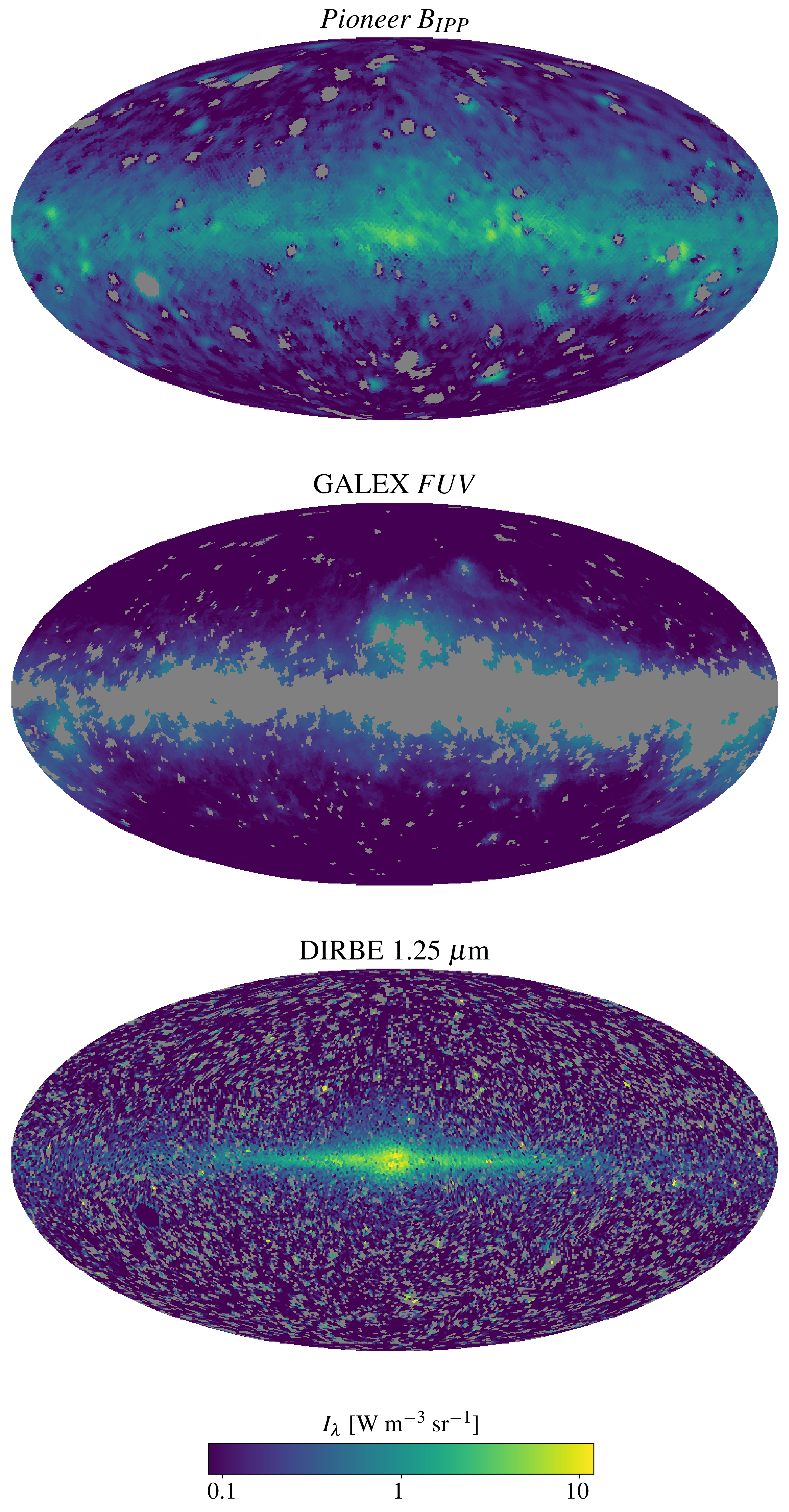}
\caption{
$I^\mathrm{diffuse}_\lambda$ derived from observations (see text for details).
}
\label{fig:res}
\end{figure}

In a first attempt to model diffuse radiation, I follow \citet{MasanaMNRAS2021} and exploit the published 
correlations to derive maps of $I_\lambda^\mathrm{diffuse}$ from those of $I_\nu(100\,\mu\mathrm{m})$. From $I_\lambda^\mathrm{diffuse}$, the diffuse component of the LISRF $J_\lambda$ is then computed.
I verify the validity of the $I_\lambda^\mathrm{diffuse}$ maps by comparison with maps 
of diffuse emission that can be derived from the observations and the direct starlight estimates
of the current work. For each of the two {\em Pioneer} and the
DIRBE 1.25~$\mu$m bands, I use the map presented in App.~\ref{app:newe} (including all emission components)
from which I subtracted a map of direct starlight, 
estimated by integrating the spectra of Sect~\ref{sect:spectrum}
over the filter bandpasses and convolving the result by the
appropriate resolution: I used FWHM = $4^\circ$
for the {\em Pioneer} data, with an area close to the average field-of-view of the IPP instruments; and 0.7$^\circ$ for DIRBE. For the 
NUV and FUV bands, I directly used the background maps derived by \citet{MurthyApJS2014} from observations taken by GALEX: they were obtained 
after masking point sources and averaging over various fields,
and cover most of the sky, but excluding the galactic plane. 
From each ot the maps descrived above,
I further subtracted the value of the extragalactic background
listed in \citet{DriverApJ2016}; however, this does not modify significantly the results. I note here that the contribution of zodiacal light is absent in the {\em Pioneer} bands, while was removed 
using a model from DIRBE and GALEX maps.
In Fig.~\ref{fig:res} I show the
maps of $I_\lambda^\mathrm{diffuse}$ for the $B_\mathrm{IPP}$, $FUV$ and 1.25~$\mu$m bands, and in Fig.~\ref{fig:latfit} the surface brightness
profiles along  Galactic latitude, averaged over the whole longitude range, for all the five bands I considered. Fig.~\ref{fig:latfit} also presents the profile for $I_\nu(100\,\mu\mathrm{m})$ from the map of \citet{SchlegelApJ1998}, after subtracting a cosmic background of 0.34 MJy sr$^{-1}$ \citep{DriverApJ2016}.

\subsection{The literature solution}

\citet{BrandtApJ2012} study residual emission at high galactic latitude in 92000 blank-sky optical spectra from the Sloan Digital Sky Survey (SDSS). The works has been recently revised by \citet{ChellewApJ2022},
by expanding the sample of SDSS blank-sky spectra to $\sim$250000, and including dust self-absorption. 
They use the 100~$\mu$m maps
from the IRIS reprocessing of IRAS data \citep{MivilleDeschenesAPJS2005}. Assuming that dust emission is due
to radiation absorbed from the LISRF, $I_\nu$(100$\mu$m) scales 
as the product 
of the LISRF intensity and the dust column density. In the case of low dust column density, $I_\nu$(100$\mu$m)
should correlate linearly with any emission from scattered starlight. As the dust optical depth increases, 
however, only a fraction of the column density contributes to scattering starlight in the direction of the observer.
This effect is taken into account
by \citet{ChellewApJ2022} by assuming
\begin{equation}
I_\lambda^\mathrm{diffuse} \propto
\beta_\lambda\times 
I_\nu(100\,\mu\mathrm{m}),
\label{eq:beta}
\end{equation}
with
\begin{equation}
\beta_\lambda=\frac{(1-e^{-\tau_\lambda})}{\tau_\lambda}
\label{eq:beta2}
\end{equation}
where the correction factor $\beta_\lambda$ depends on the optical depth 
$\tau_\lambda$ for the diffuse radiation: in the optically thin 
case, $I_\lambda^\mathrm{diffuse}$ is proportional to 
$I_\nu(100\,\mu\mathrm{m})$, while in the optically thick case
it saturates to a value independent of the line of
sight (both $\tau_\lambda$ and $I_\nu(100\,\mu\mathrm{m})$
being dependent linearly on the dust column density). 
The authors derive $\tau_\lambda$ from the 
\citet{SchlegelApJ1998} extinction maps, assuming the average
extinction law of \citet{FitzpatrickPASP1999}.

 \begin{figure*}
\centering
\includegraphics[height=5.6cm]{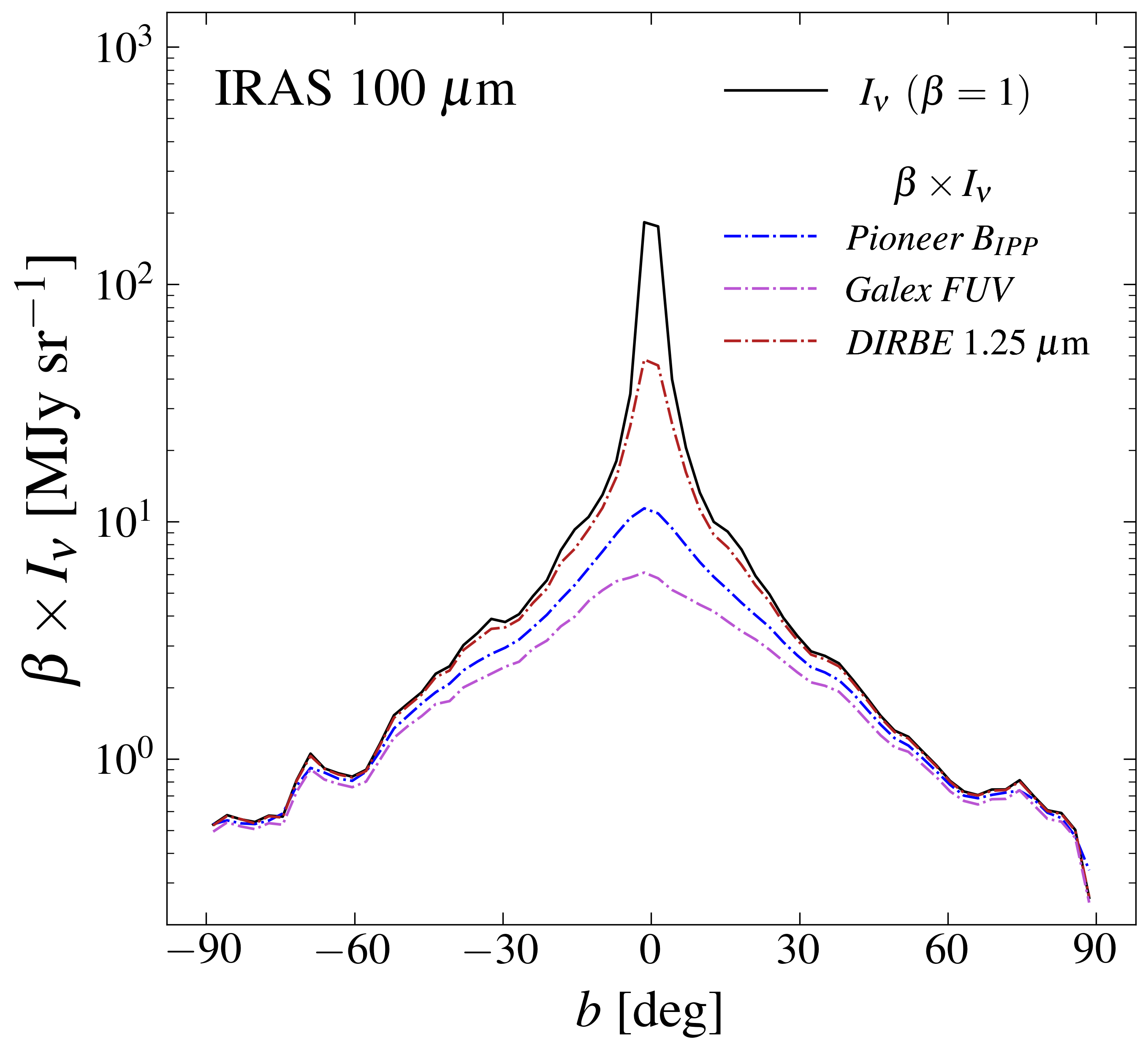}\includegraphics[height=5.6cm,trim={-1.8cm 0 0 0},clip]{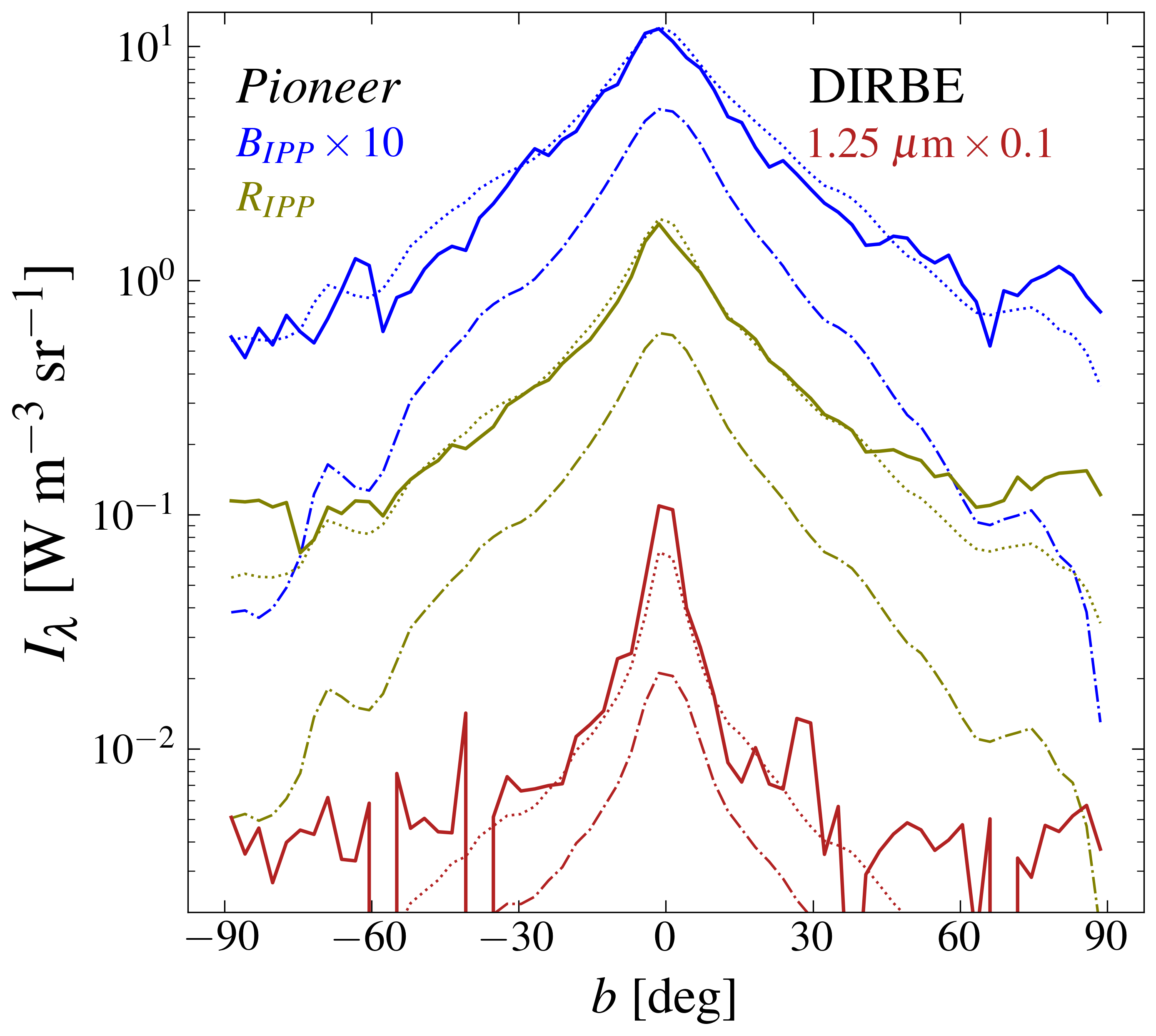}\includegraphics[height=5.6cm,trim={3.6cm 0 0 0},clip]{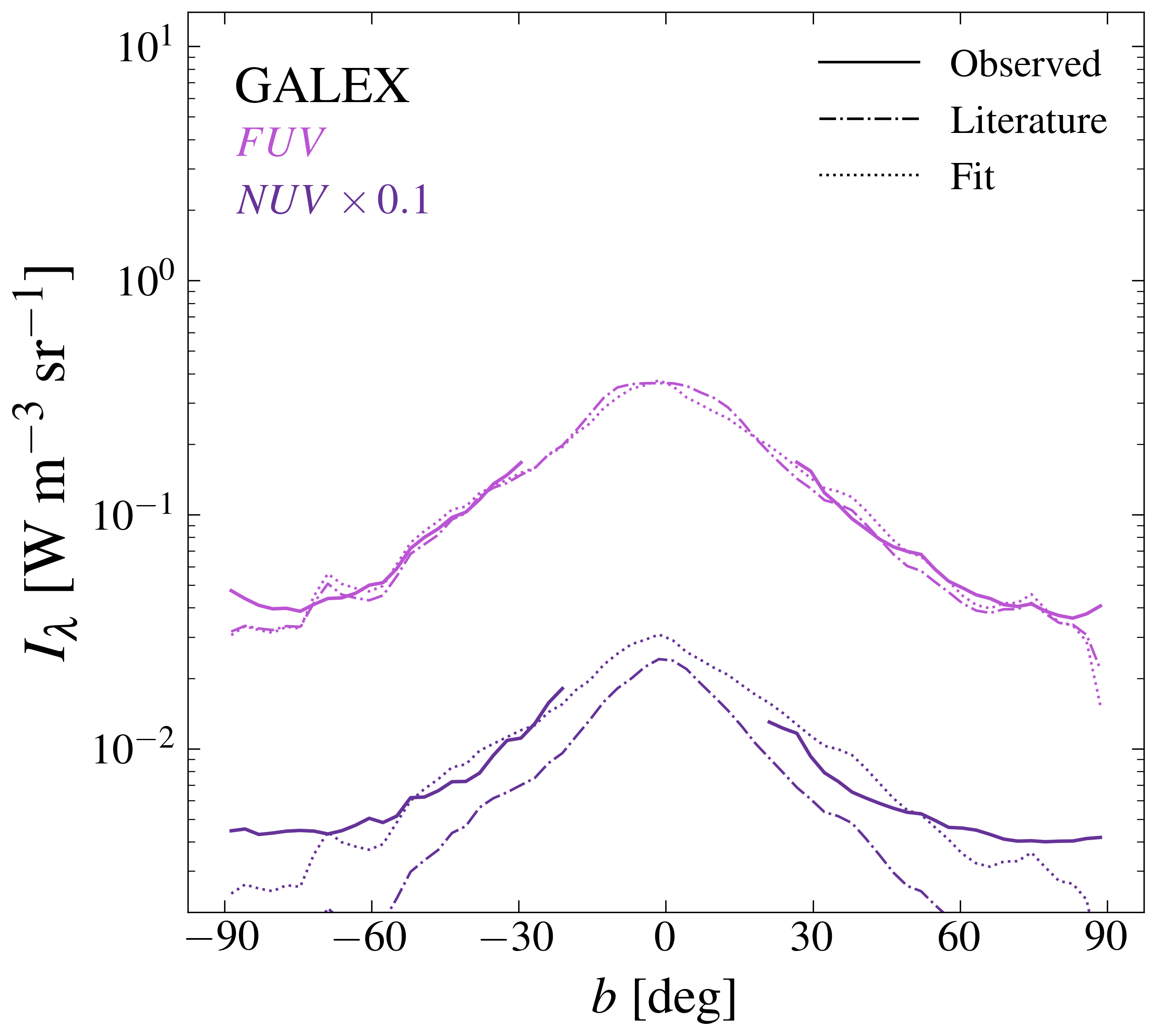}
\caption{
Galactic latitude profiles averaged over the whole longitude range. 
In the left panel I show $I_\nu(100\,\mu\mathrm{m})$ and the
$\beta \times I_\nu(100\,\mu\mathrm{m})$ templates for the $B_{IPP}$,
$FUV$ and 1.25~$\mu$m bands. In the two panels on the right
I show the profiles for diffuse radiation
directly estimated from observations and those from the {\em literature} and {\em fit} solutions (see text for details).
}
\label{fig:latfit}
\end{figure*}

\begin{figure*}
\centering
\includegraphics[width=\hsize]{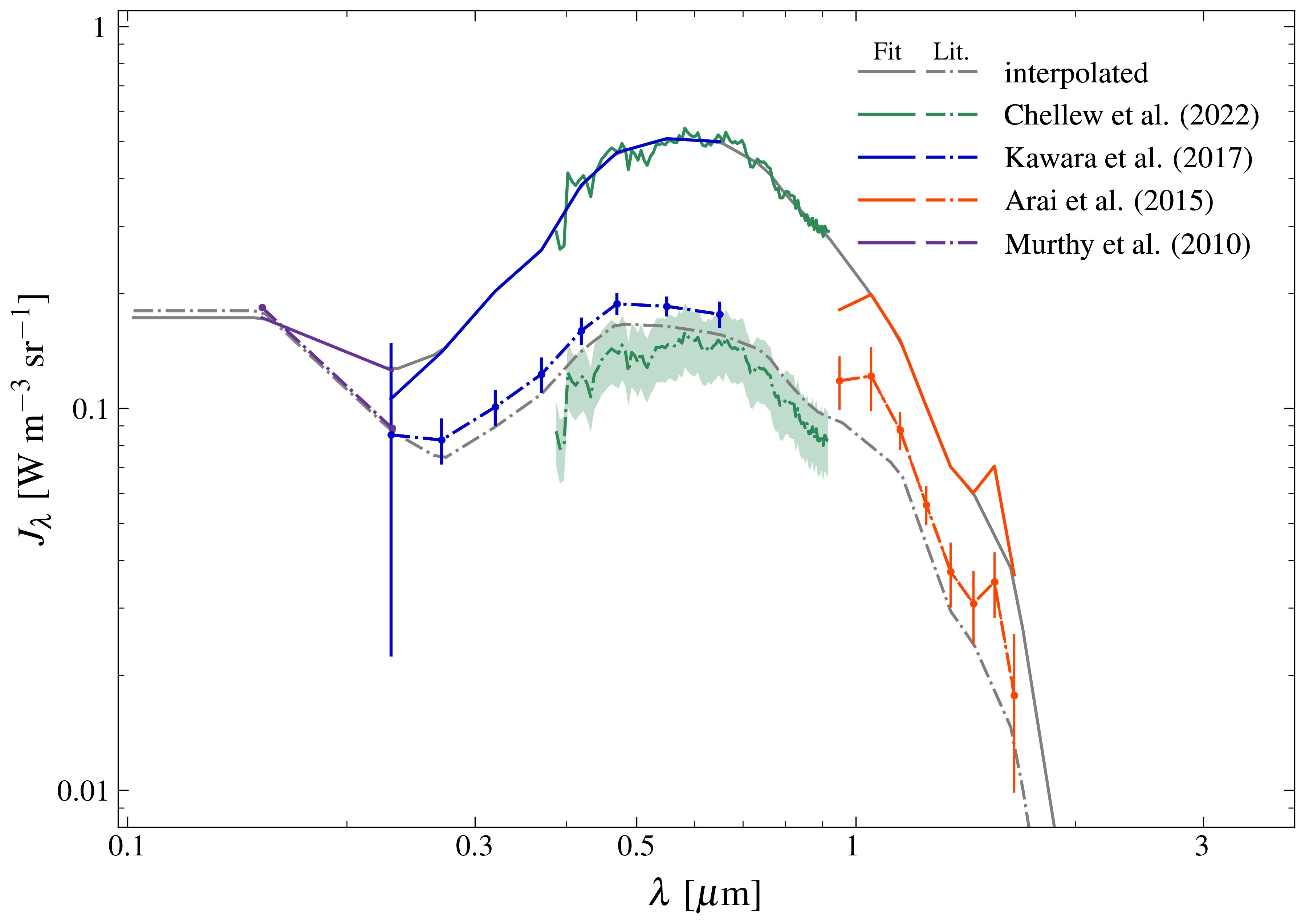}
\caption{
The diffuse starlight component of the LISRF spectrum. I show the {\em literature} and {\em fit} estimates together with the interpolated solutions (see text for details).
}
\label{fig:LISRFd}
\end{figure*}

Using the same data and the results from \citet{ChellewApJ2022}, I derive a map for $I_\lambda^\mathrm{diffuse}$ from the
IRIS map $I_\nu$(100$\mu$m), over the 
wavelength range analyzed by the authors (0.4 $\la \lambda/\mu\mathrm{m} \la$ 0.9) and the entire sky.
The analyses of \citet{BrandtApJ2012}
and \citet{ChellewApJ2022} suffer from
a bias of uncertain origin, which can
be corrected by a multiplicative factor. I 
also scaled the $I_\lambda^\mathrm{diffuse}$ maps for
this factor, and derive $J_\lambda$; it is shown in
Fig.~\ref{fig:LISRFd}, together with the larger 
source of uncertainty estimated by the authors, that
on the bias factor itself. I label this and other estimates 
directly based on published correlations as {\em literature} solution. There are clearly other uncertainties that are
difficult to estimate, since I applied to the entire sky the
results of \citet{ChellewApJ2022},
which are based on data at high Galactic latitude and mainly around the Galactic North Pole. In particular, I estimate $I_\lambda^\mathrm{diffuse}$ also on the Galactic plane, where the emission
traced by $I_\nu$(100$\mu$m) in specific regions might not
be directly related to the dust column 
density, because of dust heating by 
localized star-forming regions and not by the average LISRF. 

\citet{KawaraPASJ2017} adopted a similar approach to \citet{BrandtApJ2012} and fit emission in blank-sky spectra 
from the FOS instrument aboard the Hubble Space Telescope
with a model including diffuse starlight and zodiacal light.
Diffuse starlight is modelled using the
$I_\nu$(100$\mu$m) map from \citet{SchlegelApJ1998}, after subtracting an extragalactic background of 0.8 MJy sr$^{-1}$.
The sky coverage is smaller than in \citet{BrandtApJ2012} and \citet{ChellewApJ2022}, being limited to just 54 pointings, but it includes data at low Galactic latitude. 
The model includes a linear dependence of $I_\lambda^\mathrm{diffuse}$ on $I_\nu$(100$\mu$m),
for the optically thin regime, and
a negative term depending on the square of $I_\nu$(100$\mu$m), to take into account the saturation at high optical depths (see also \citealt{IenakaApJ2013}). 
I derive $I_\lambda$ and  $J_\lambda$ from the data and analysis of \citet{KawaraPASJ2017} for
0.2 $\la \lambda/\mu\mathrm{m} \la$ 0.7. Since the formulation chosen by \citet{KawaraPASJ2017} cannot be applied beyond the $I_\nu$(100$\mu$m) range they studied (otherwise the
quadratic term prevails and the diffuse emission becomes negative), I assume that the correlation saturates for $I_\nu$(100$\mu$m)$\ge 50$~MJy  sr$^{-1}$ (corresponding to a visual extinction of $\sim$5; \citealt{IenakaApJ2013,ChellewApJ2022}). The corresponding $J_\lambda$
in shown in Fig.~\ref{fig:LISRFd}, together with the errors from the correlation estimated by the authors.

\citet{AraiApJ2015} studied the diffuse radiation in NIR 
spectra taken in a few moderate-to-high Galactic latitude fields,
by using the LRS instrument aboard the sounding-rocket borne Cosmic Infrared Background Experiment. They assume a simple linear relation
between diffuse starlight and $I_\nu$(100$\mu$m) (from \citealt{SchlegelApJ1998}) since in their
fields, and at the wavelength they explore (0.95 $\le \lambda/\mu\mathrm{m} \le$ 1.65) the opacity is small. They also
include zodiacal light in the modelling.
As for the previous works, I
use their results assuming that the ratio between scattered light and FIR radiation saturates for $I_\nu$(100$\mu$m)$\ge 50$~MJy  sr$^{-1}$ (where extinction becomes $\ga1$), thus mimicking 
the change from the optically thin to the optically thick regime (Fig.~\ref{fig:LISRFd}).

The same approach is adopted when deriving $J_\lambda$ in the UV 
(Fig.~\ref{fig:LISRFd}). I use the linear correlation found by 
\citet{MurthyApJ2010} from the background FUV and NUV GALEX maps already discussed.
In this case I impose saturation at $I_\nu$(100$\mu$m) $ \ge 10$~MJy  sr$^{-1}$, when the
medium becomes optically thick in the GALEX bands \citep{MurthyApJ2010}.

The wide span of methods, sky coverage and dynamical ranges used in the works described in this
Section reflects into the piece-wise aspect of the spectrum in Fig.~\ref{fig:LISRFd}. In order 
to have a unique {\em literature} solution over the whole UV-to-NIR wavelength range,
I (rather arbitrarily) constructed an interpolated $J_\lambda$ spectrum
passing in between the various estimates. Beyond the
wavelength ranges studied in the literature, I assume that the 
spectrum of diffuse light is constant for $\lambda\le 0.15\,\mu$m, and goes to zero for $\lambda> 2\,\mu$m. The spectra derived 
from the various works are within 25\% of the interpolated solution.
I finally produced $I_\lambda^\mathrm{diffuse}$ maps from 
the various results and scale them to match the
interpolated spectrum.

In Fig.~\ref{fig:latfit} I plot the Galactic latitude profiles
of the $I_\lambda^\mathrm{diffuse}$ maps from the {\em literature}
solution, for the bands at which I derived observation-based
estimates. Apparently, the {\em literature} solution is not able 
to match the observations in the optical and NIR.  
Instead, the agreement between model and observations is good 
in the FUV (right panel) and also in the NUV, if 
a zero-level emission, of unknown  origin but uncorrelated to dust \citep{MurthyJAA2023}, is subtracted. 
I will discuss these issues later.

\subsection{The fit solution}

As an alternative solution, I construct a template for $I_\lambda^\mathrm{diffuse}$ using Eq.~\ref{eq:beta}, the 
\citet{SchlegelApJ1998} maps for extinction and 
$I_\nu$(100~$\mu$m), and the \citet{FitzpatrickPASP1999} average extinction law. The effects of the $\beta_\lambda$  corrective term (Eq.~\ref{eq:beta2}) is shown in the left panel of Fig.~\ref{fig:LISRF}, for a few representative bands: at
shorter $\lambda$'s, the Galactic latitude profile of the
template becomes shallower
and can describe better the shallow 
trend of $I_\lambda^\mathrm{diffuse}$ derived from observations, 
without the need of introducing a saturation cut in $I_\nu$(100~$\mu$m). Thus, the template can be scaled to match, 
on average, the maps of diffuse emission derived from 
observations, by using  a single scaling factor
over the whole dynamic range of $I_\nu$(100$\mu$m).

The comparison between the observed profiles and those
for the rescaled templates is shown in the right panels of Fig.~\ref{fig:latfit}. Despite the crudeness of the modelling,
there is a satisfactory agreement in the {\em Pioneer} bands. 
In the FUV, $I_\lambda^\mathrm{diffuse}$ is very close to the
{\em literature} estimate. The result in the NUV would be similar 
if I had subtracted the large zero-level discussed earlier; however, in order to retain all emission, the template is scaled 
to the full NUV emission. The worse agreement is in the DIRBE 1.25~$\mu$m band, despite the analysis is restricted to the
galactic plane ($b<40^\circ$) to avoid noisy regions. I discuss
these issues in the next Section.

Finally, I use the scaling factors found for the bands of Fig.~\ref{fig:latfit} to rescale the published spectra.
In analogy with what done in the previous section, 
I call the interpolation over the whole UV-optical-NIR
range the {\em fit} solution: it is shown in Fig.~\ref{fig:LISRFd}.
By construction, the $J_\lambda$ spectrum in the optical-NIR
is higher for the {\em fit} solution, by about a factor 3.
I will show in the next Section the impact of the two different solutions on the total LISRF spectrum.

\begin{figure*}
\centering
\includegraphics[width=\hsize]{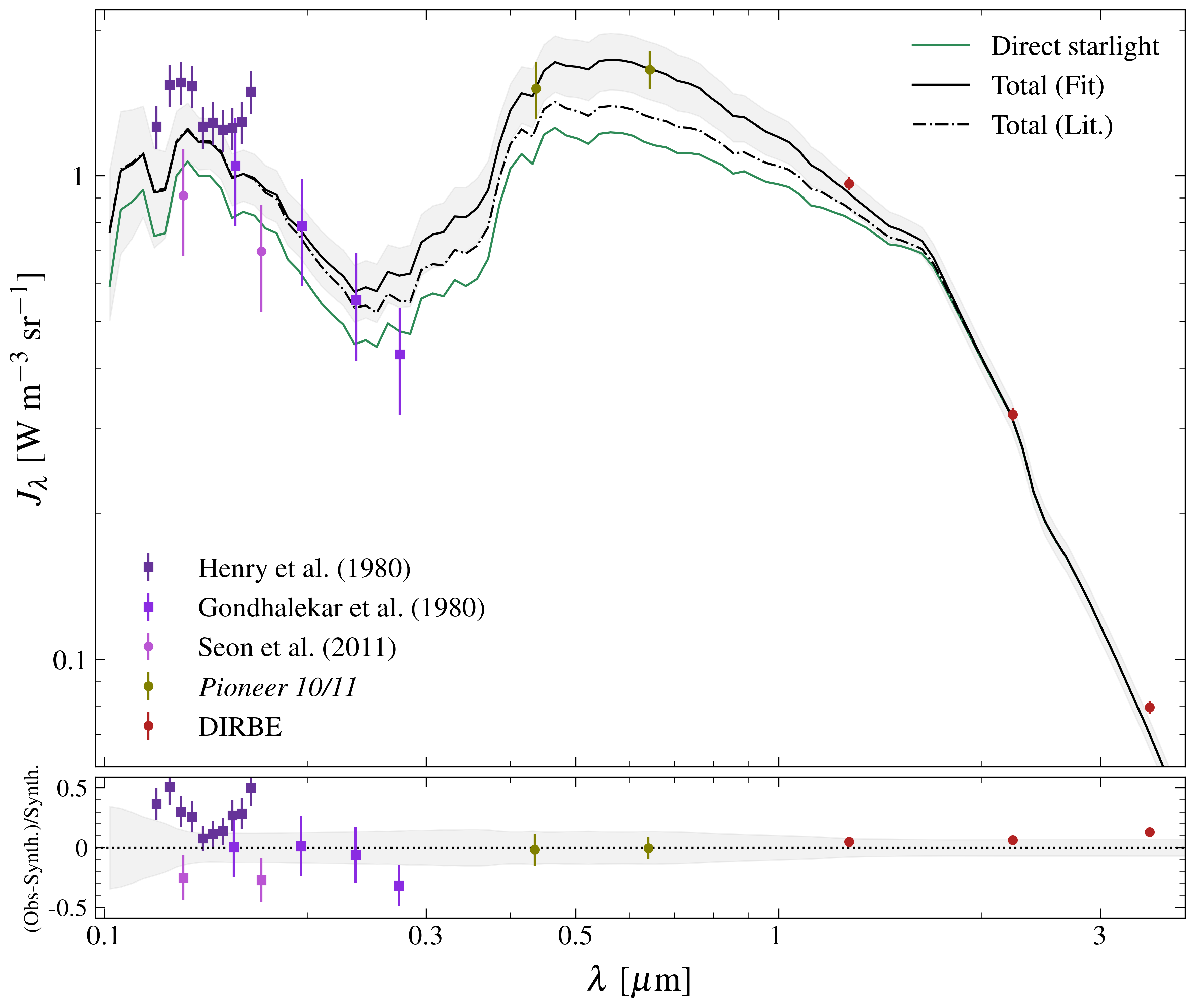}
\caption{
The total LISRF spectrum compared to estimates derived from large-area observations. The bottom panel show the relative difference between observations and synthetic photometry derived from the {\em fit} solution spectrum.
}
\label{fig:LISRFa}
\end{figure*}

\section{The total LISRF spectrum: results and discussion}
\label{sect:total}

I added the LISRF spectrum from direct starlight (Sect.~\ref{sect:spectrum}) to the {\em literature} and {\em fit} solutions for the spectrum of diffuse light (Sect.~\ref{sect:diffuse})
to obtain two estimates of the total LISRF: they are shown in Fig.~\ref{fig:LISRFa}. The two estimates are compared
with those values of  $J_\lambda$ presented in 
Sect.~\ref{sect:compa} that are obtained
(or estimated) on a large fraction of the sky.

As discussed in the previous section, diffuse radiation estimated from the
{\em literature} solution, when added to direct starlight, is not able to reproduce the independent 
estimates of $J_\lambda$ in the optical and NIR. The {\em fit} solution instead is made 
to pass through (or near) them by construction. For the {\em Pioneer} bands, a possible reason 
for the apparent failure of the {\em literature} solution could be that the estimates from 
observations are biased high. As discussed in Appendix~\ref{app:pioneer}, the contribution of
bright stars was removed from the distributed {\em Pioneer} data; yet the data might  
still includes a significant residual contribution from those objects. 
As a result, a fraction of direct starlight could have been counted {\em twice}, being 
included in both the recovered emission for $V<6.5$ stars and 
possibly in the available maps including stars with $V>6.5$
and diffuse radiation (see Appendix~\ref{app:pioneer} for
details). This uncertainty is included in the large errors
shown in Fig.~\ref{fig:LISRFa}. The {\em literature} solution could 
be consistent (though marginally) with the $B_\mathrm{IPP}$ datapoint. However,
$J_\lambda$ in $R_\mathrm{IPP}$ still requires a contribution from 
diffuse starlight that the {\em literature} solution cannot provide.

In the DIRBE 1.25 $\mu$m band, the total $J_\lambda$ is not only higher than the 
{\em literature} solution, but also of the {\em fit} one. In fact, the template for 
diffuse starlight does not match the estimate based on observations near the
galactic center (see the excess emission on the plane in Fig.~\ref{fig:latfit}). The reason is shown by the 
map of Fig.~\ref{fig:res}: a central structure can be seen, 
resembling the Galactic bulge (obviously missing from the 
IRAS-based template). 
Clearly, the model of Sect.~\ref{sect:spectrum} was not sufficient 
to remove the full direct starlight contribution. This could be
the results of highly extinguished stars emitting predominantly
in the NIR band (but missing from the {\em Gaia} catalog, as discussed in Sect.~\ref{sect:spectrum}); or
to the inaccuracy of the approximation for the spectrum of
the redder {\em Gaia} stars without APs. The same issue could 
be the cause for the flat (but noisy) background at high galactic latitude (see Fig.~\ref{fig:latfit}). In any case, once the
{\em fit} solution is subtracted, the residual emission 
accounts only for $J_\lambda\approx0.03$ W m$^{-3}$ sr$^{-1}$ (mostly from the
bulge region with $|b| < 30^\circ$, $|l| < 30^\circ$).
This value is of the order of the error on the full 
$J_\lambda$ estimate in this band (App.~\ref{app:dirbe}), and
about half of the uncertainty estimated in Sect.~\ref{sect:spectrum}
for the model of diffuse starlight. Thus, I can still
consider the {\em fit} solution consistent with observations.

Another reason for the failure of the {\em literature} solution
might reside in the smaller dynamic range and spatial coverage of 
the data used to derive the correlations derived in the optical and NIR. The bulk of the 
contribution of diffuse starlight to $J_\lambda$ comes from
low Galactic latitude ($\approx 70\%$ from $|b|<20^\circ$
in the optical and NIR), while most of the analysis in the 
literature is done on high Galactic latitude regions, with the
exception of \citet{KawaraPASJ2017}, which do include observation close to the Galactic plane. Also, 
while here templates are derived
after averaging the observations over pixels of $\sim 1^\circ\times1^\circ$, the individual observations of \citet{KawaraPASJ2017}
and \citet{ChellewApJ2022} have field-of-views of few arcsec squared;
those of \citet{AraiApJ2015} do cover field-of-views of a few degrees squared (though not continuously, but with a series of thin slits)
but are limited to just six high latitude regions. Thus, one
might wonder if those correlations describe better variation with
$I_\nu$(100$\mu$m) at smaller scales than the single template 
scaled on the whole sky.

The variety of methods also might cause some 
differences. For example, only \citet{KawaraPASJ2017} subtract
the extragalactic background (though using a value higher than
what assumed here, thus causing the steeper latitude gradients 
of the {\em literature} profiles in the optical shown in Fig.~\ref{fig:latfit}). \citet{ChellewApJ2022} do 
not explicitly consider the impact of zodiacal light; also, 
their results have to be scaled by a bias factor in order to 
reproduce the results of a radiative transfer model of diffuse 
starlight \citep{BrandtApJ2012}. Even after applying the correction,
the results (and thus the radiative transfer model used to rescale 
them) are still offset from those of \citet{KawaraPASJ2017}
and \citet{AraiApJ2015} (see also Fig.~\ref{fig:LISRFd}).

The difference between the two solutions is negligible in the 
UV, where the correlations from \citet{MurthyApJ2010} 
was derived on a large fraction of the sky: their results, coupled
with the simple model for the saturation with dust extinction, 
are very similar to the current template-based approach, at least in the FUV, but also in the NUV if the zero-level offset found by
\citet{MurthyApJS2014} had been removed. 

In the following, I adopt the total $J_\lambda$ including the {\em fit} solution as the correct LISRF representation. Since the spectrum is made to pass through the {\em Pioneer} datapoints, I require the total 
uncertainty of the spectrum to match the uncertainty from those
observations (for a conservative estimate, I use the larger error for $B_\mathrm{IPP}$): this can be achieve by assuming a relative uncertainty
of $\sim40\%$ for diffuse starlight (added in quadrature to the uncertainty for direct starlight presented in Sect.~\ref{sect:spectrum}).
The same relative error is used for the whole wavelength range and the
total is shown with the shaded area in Fig.~\ref{fig:LISRFa}. The total
uncertainty in the optical range is 13\% (again by construction). In the
UV (in the range of the $FUV$ and $NUV$ bands) the uncertainty rises to 16\% (because of the higher uncertainty of direct starlight, while the contribution of diffuse starlight is reduced with respect to the optical). This value is lower than the average scatter of the UV datapoints around the {\em fit} solution, $\sim25\%$, probably due
the large calibration uncertainties of the observations. Longer wavelengths, instead, are dominated by direct starlight and 
the uncertainty is $\sim7\%$, as estimated in Sect.~\ref{sect:spectrum}; the larger discrepancy of the DIRBE 3.5 $\mu$m datapoints might come from the additional contribution of dust in emission (not considered in this analysis). However, radiation in this part of the spectrum constitutes only a minor part of the total energy of the LISRF.

\begin{figure}
\includegraphics[width=\hsize]{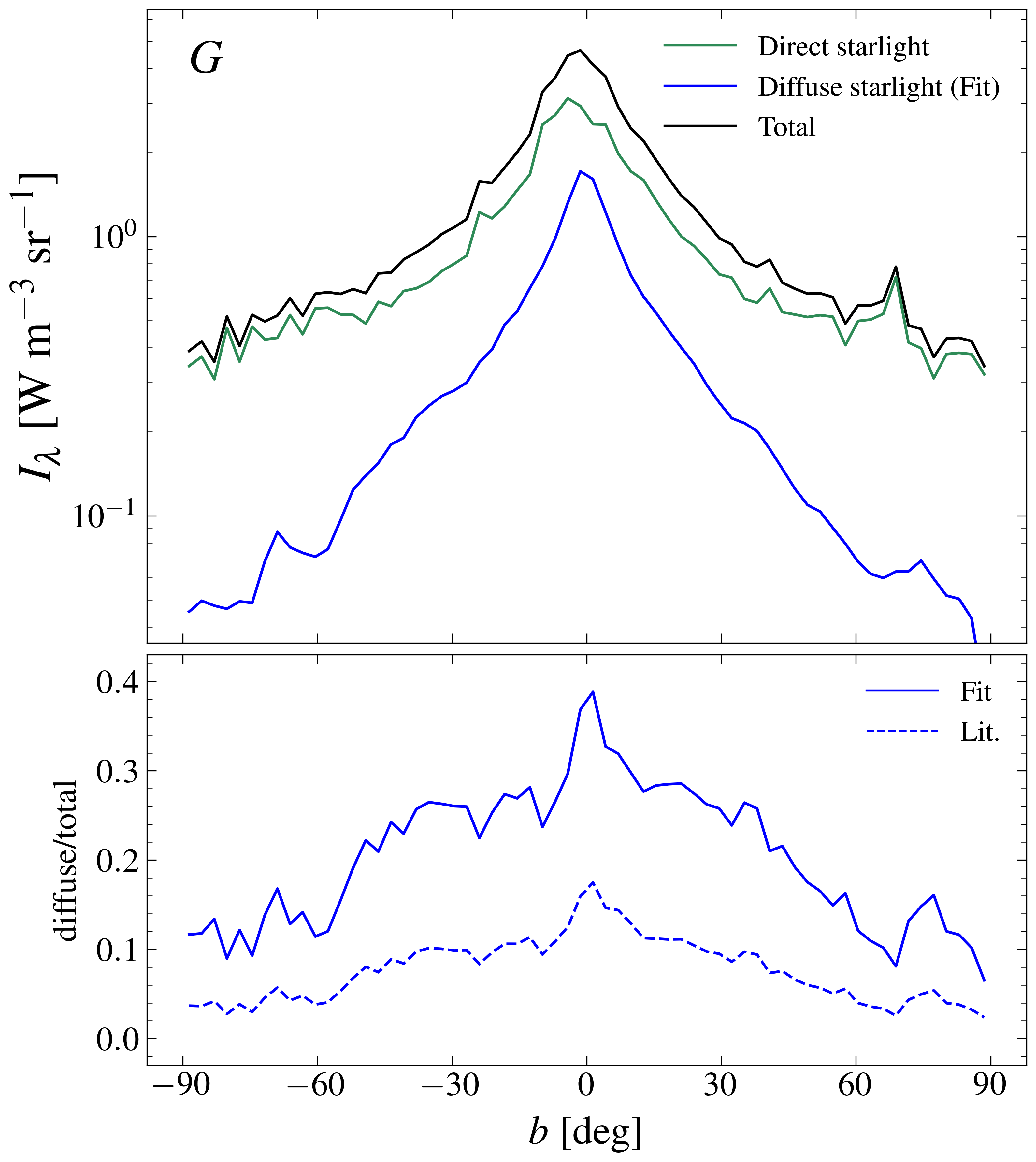}
\caption{
Galactic latitude profile of $I_\lambda$, predicted for the {\em Gaia} $G$ band, for direct and diffuse starlight, and the total ({\em fit} solution). The lower panel shows the fractional contribution of diffuse starlight for both {\em fit} and {\em literature} solutions.
}
\label{fig:latfrac}
\end{figure}

Diffuse starlight contributes on average to 
$\approx20\%$ of the total LISRF in the UV for $\lambda<0.3\,\mu$m, rising to $\approx$30\% in the optical for $0.4\,\mu\mathrm{m}<\lambda<0.7\,\mu$m, then falling in the NIR, with a
10\% contribution at 1.25~$\mu$m
(for the {\em fit} solution; $\approx$15, 10 and 5\%, respectively, for the {\em literature} solution). In Fig.~\ref{fig:latfrac} I
show the  $I_\lambda$ profile predicted for the {\em Gaia G} band
and the variation of the fraction of diffuse starlight with latitude:
beside a central peak, the contribution of diffuse starlight is relatively constant ($\approx 25\%$) with latitude up to $|b|\approx 50^\circ$, dropping to 10\% towards the poles (for the {\em fit} solution).

\begin{figure*}
\centering
\includegraphics[height=7cm,trim={0 3.5cm 0 0},clip]{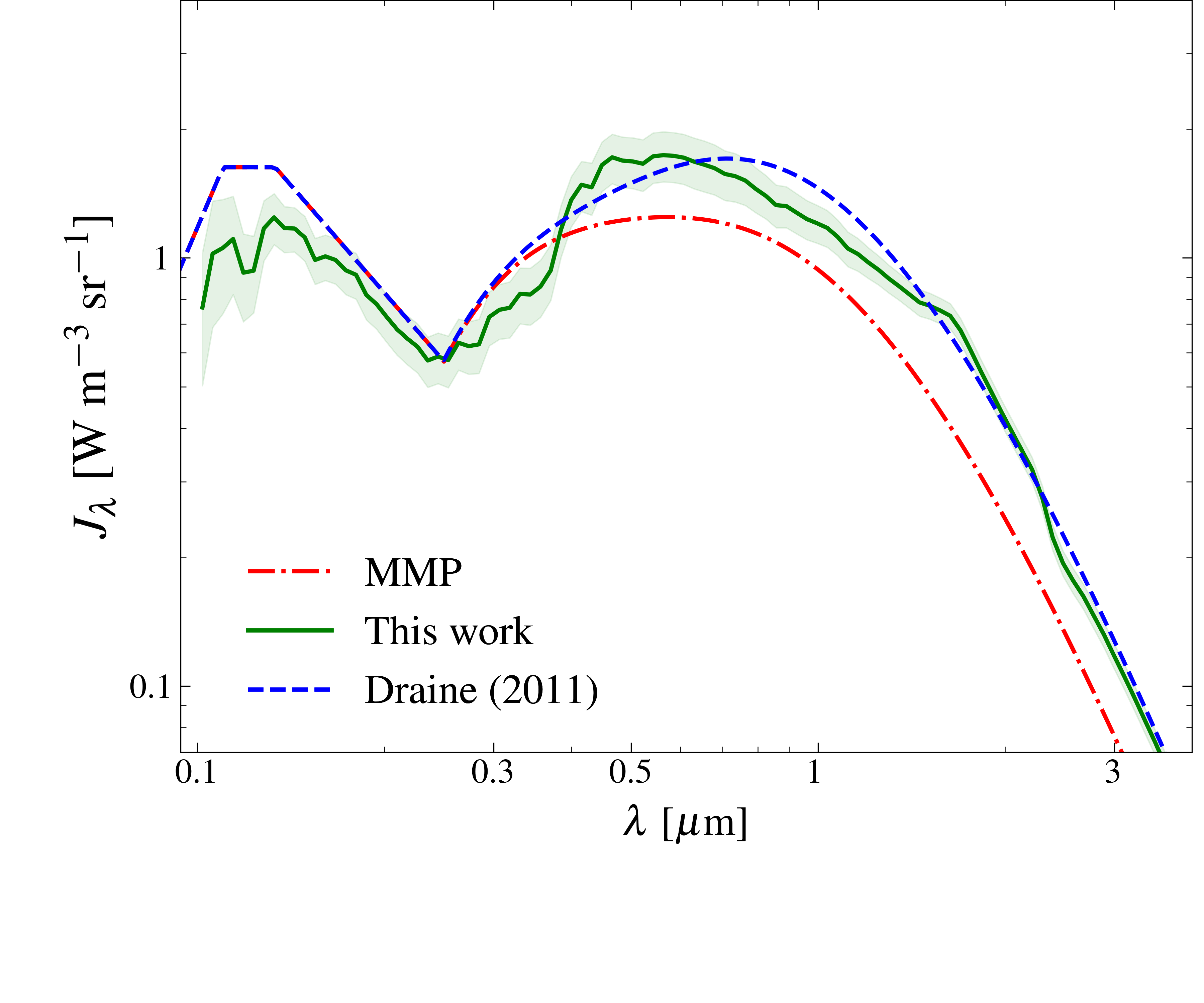}\includegraphics[height=7cm,trim={4.4cm 3.5cm 0 0},clip]
{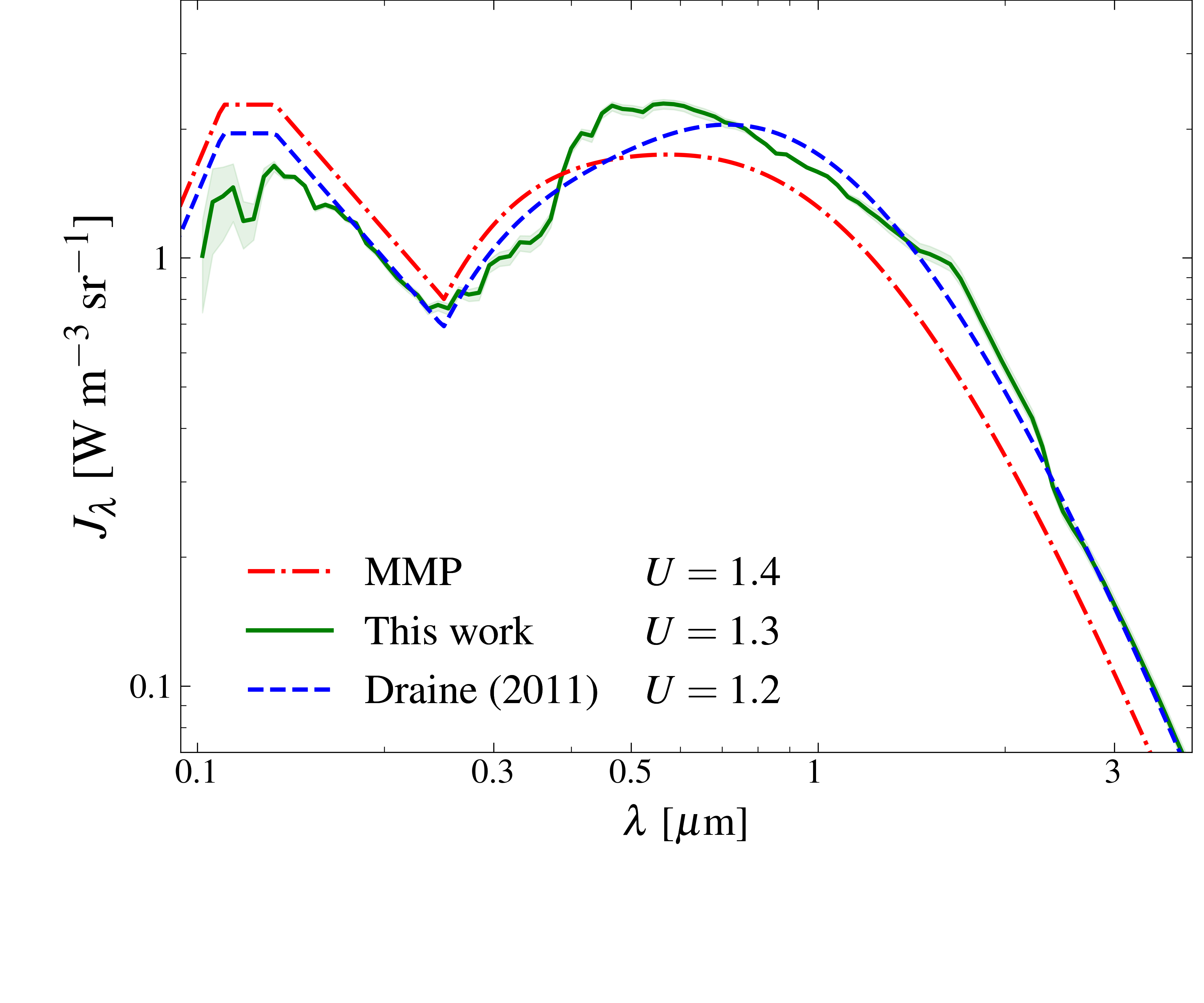}
\includegraphics[height=7cm,trim={0 3.5cm 0 0},clip]{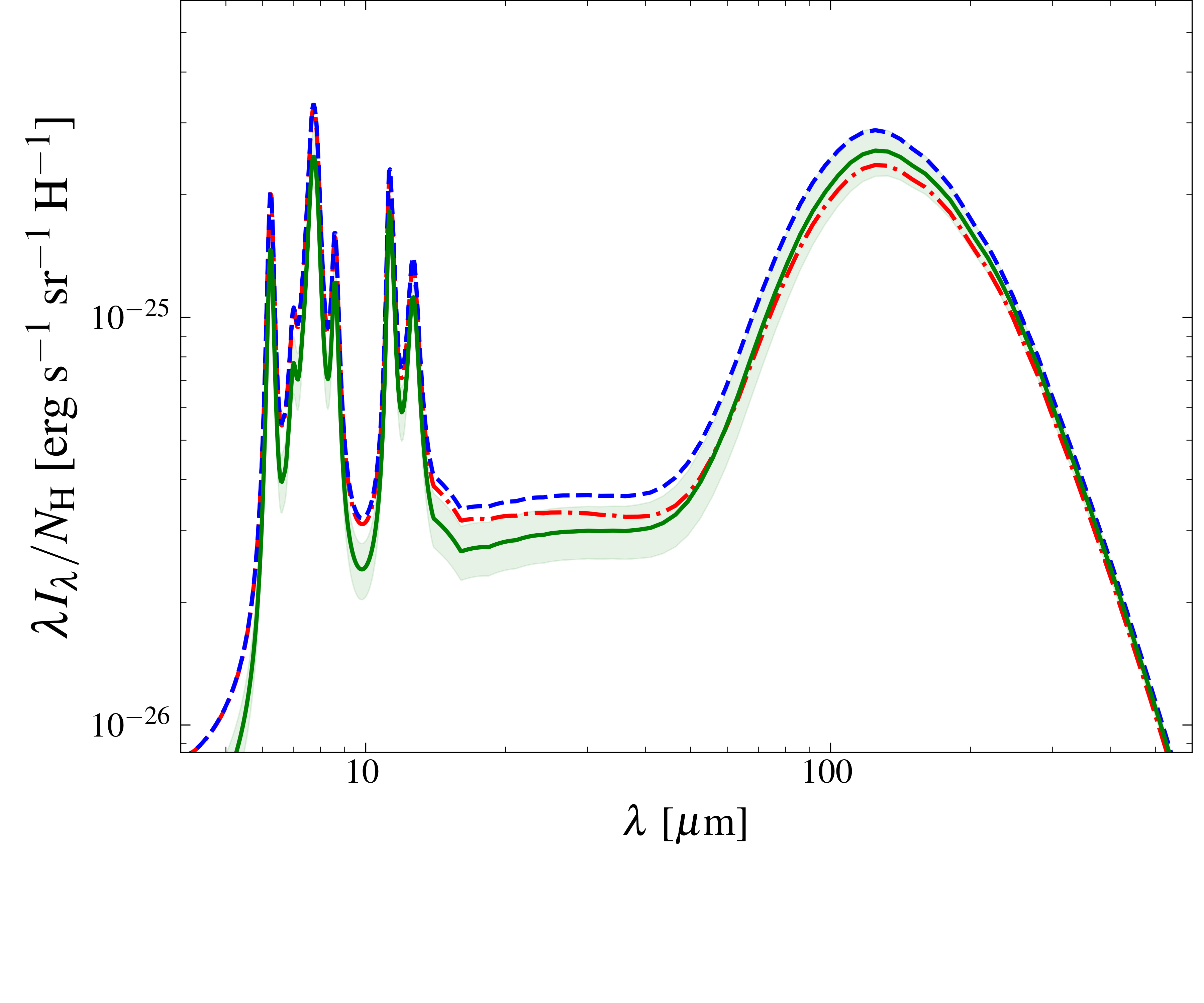}\includegraphics[height=7cm,trim={4.4cm 3.5cm 0 0},clip]{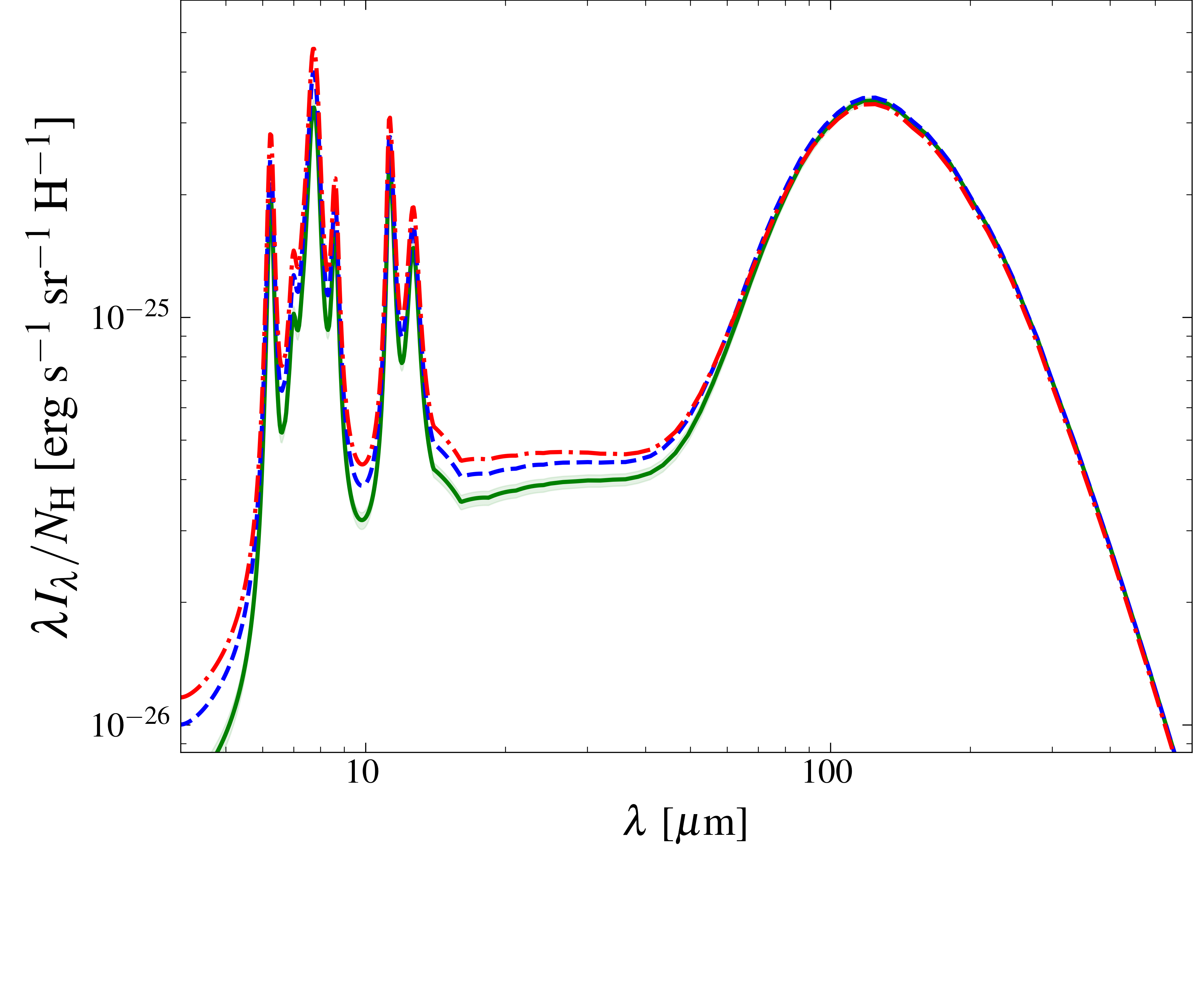}
\caption{
Top row: the left panel shows the LISRF from this work ({\em fit} solution, with its uncertainty), MMP and \citet{DraineBook2011}; the righ panel shows the same, but LISRFs are rescaled by the factor $U$ in the legend. Bottom row: Dust emissivity predicted for the \citet{JonesA&A2017} dust model under different heating conditions. Each bottom panel shows the results for 
the LISRFs presented in the corresponding upper panel.
}
\label{fig:dustem}
\end{figure*}

\begin{table}
\caption{Selected in-band values of the new LISRF vs MMP.}
\label{tab:Jcompa}
\centering
\begin{tabular}{lcc}
                &\multicolumn{2}{c}{$J_\lambda$ [W~m$^{-3}$ sr$^{-1}$]}\\ \\
                & this work     & MMP\\ \hline \\
$FUV$           & $1.05\pm0.13$ & $1.29\pm0.19$ \\
$NUV$           & $0.64\pm0.08$ & $0.70\pm0.10$ \\
$G_\mathrm{BP}$ & $1.63\pm0.22$ & $1.22\pm0.18$ \\
$G$             & $1.55\pm0.20$ & $1.18\pm0.18$ \\
$G_\mathrm{RP}$ & $1.45\pm0.18$ & $1.13\pm0.17$ \\
$I$             & $1.34\pm0.15$ & $1.06\pm0.16$ \\       
$J$             & $0.94\pm0.08$ & $0.69\pm0.10$ \\
$H$             & $0.70\pm0.05$ & $0.39\pm0.06$ \\
$K$             & $0.34\pm0.02$ & $0.20\pm0.03$ \\
\hline
\end{tabular}
\end{table}

\subsection{The new LISRF vs MMP}

The new LISRF spectrum is compared with MMP in the top right panel of Fig.~\ref{fig:dustem}, while in Table~\ref{tab:Jcompa} values
of $J_\lambda$ are listed in some representative bands across the UV, optical and NIR spectrum. The differences between the two spectra are significant: 
the new LISRF is $\sim$20\% lower in FUV, and 30\%, 35\% and up to
$\sim$70\% higher in $G$, $J$ and $K$, corresponding to
discrepancies from 2 to $6\times$ the  15\% uncertainty claimed by \citet{MathisA&A1983}.

The total flux from the integration of the new spectrum over the whole wavelength range is $\sim$30 higher than in MMP. Also, the balance between UV and optical-NIR radiation is different: radiation for $\lambda < 0.4\,\mu$m constitutes $\sim$10\% of the total energy output for the new LISRF, while it is $\sim$20\% for MMP. While differences are expected in the optical and NIR, due to the new data not available at the time of the MMP papers, the larger UV output is due to the choice of \citet{MezgerA&A1982} of raising the \citet{GondhalekarA&A1980} fluxes by 15\%, in order to have a better agreement with the data from \citet{HenryApJ1980}. On the contrary, the current spectrum estimate agrees better with the original \citet{GondhalekarA&A1980} values.

Differences between the new LISRF and the model of \citet{DraineBook2011}
are reduced but still significative in the UV (where the model is the same as MMP) and at $\lambda\approx 1\,\mu$m. The total integrated flux of \citet{DraineBook2011} is slightly higher than the current estimate (and $\sim$40\%  higher than MMP), while the contribution to the total of UV radiation at $\lambda < 0.4\,\mu$m is $\sim 15\%$.

\section{An application of the new ISRF to dust heating}
\label{sect:dustem}

Models for MW dust grains are required to reproduce several observational constraints: one of them is the emissivity (surface brightness per gas column density) of high Galactic-latitude dust (the {\em cirrus}) that have been observed from the MIR to the submm by a variety of instruments (for the latest observational constraints see, e.g., \citealt{HensleyApJ2021,YsardA&A2024}). The most common approach is to assume that the {\em cirrus} is heated by the LISRF: for example, several authors have used the MMP model to validate grain mixtures proposed for dust in the diffuse ISM (like, e.g., 
\citealt{DwekApJ1997,LiApJ2001,ZubkoApJS2004,CompiegneA&A2011,SiebenmorgenA&A2014,JonesA&A2017,SiebenmorgenA&A2023}).

Because of the differences described in the previous section, one should expect that a dust grain mixture will behave differently when heated by each of the LISRFs considered here. 
I evaluate the dependence of dust emission on the LISRF assumption by using The Heterogeneous dust Evolution Model for Interstellar Solids 
(THEMIS\footnote{https://www.ias.u-psud.fr/themis}) and computing the MIR-to-submm emissivities with 
the DustEM software\footnote{https://www.ias.u-psud.fr/DUSTEM} \citep{CompiegneA&A2011}. 
The THEMIS model for dust in the diffuse ISM essentially consists of three grain populations:
i) hydrogenated amorphous carbon, a-C(:H), grains of large size (typical radius $a\approx200$ nm), with a photo-processed a-C mantle, responsible for extinction/absorption in the optical 
and making most of the FIR emission at thermal equilibrium;
ii) amorphous silicate grains with iron inclusions and a-C mantles, of similar size and contributing to the FIR emission;
iii) amorphous carbon grains (a-C) of small size (from 100 down to 0.4 nm), responsible for the rise of extinction for UV radiation end emitting stochastically in the MIR, both in the continuum (I use here $\lambda=30\,\mu$m as a reference) and in the emission bands for
$3\la\lambda/\mu\mathrm{m}\la 15$ (other models prefer to use the Polycyclic Aromatic Hydrocarbon macro-molecules, PAHs, as the carriers of these features; \citealt{DraineBook2011}). For
full details on THEMIS, see \citet{JonesA&A2013}; here I use the version documented in \citet{JonesA&A2017}.

Fig.~\ref{fig:dustem} (bottom left panel) shows the emissivity predicted for THEMIS. Overall, the dust model absorbs (and re-emits)
6\% more radiation when heated by the LISRF derived in this work
(and 17\% for \citealt{DraineBook2011}) with respect to when MMP
is used. The ratios are different with respect to those for the LISRF total fluxes, because of the different spectra of the LISRFs and of 
the wavelength-dependent absorption and scattering properties of THEMIS (set to match the observational constraints of the average MW extinction law and albedo). At the
peak of thermal emission, the emissivity for the newer LISRF is 8\% higher
than for MMP; 9\% smaller at $\lambda=30\,\mu$m, where the emission
is due to stochastically heated small grains; and up to 24\% smaller within the MIR bands. 
For the \citet{DraineBook2011} LISRF
the differences are larger at thermal peak (22\% more) and reducing when going to shorter wavelengths up to the MIR features (from 11 to just 2\% more, accounting for the similar UV spectrum). In the bottom left panel of Fig.~\ref{fig:dustem}, I also show with a shaded area the spread in emissivities delimited by the DustEM output obtained when the LISRFs are the lower and upper boundaries of the uncertainties shown in the top left panel. Most of the differences between the various LISRF are within the shaded area, with the notable exception of the MIR bands.

Since THEMIS meets the observational constraints with MMP, the red curve in the bottom-left panel of Fig.~\ref{fig:dustem} is representative of the observed emissivity.
If \citet{JonesA&A2017} had used one of the other estimates for the
LISRF, they would have needed to change the properties of the dust 
model (so that the blue/green curves match the red one). For
example, when using the LISRF from this work one would need to lower 
the contribution of large grains responsible for the thermal peak and raise that of the smaller grains dominating the MIR emissivity. For the same grain size distribution and material composition, this could be naively obtained by changing the dust-to-gas ratios of these components, provided that all other observational constraints are met (and not only emissivity).

For their new model, \citet{HensleyApJ2023} rescale the \citet{DraineBook2011} LISRF by 
multiplying it by a constant factor $U=1.6$ at all wavelengths. The rescaling was required in order to have 
consistency between the energy absorbed by dust in the UV-to-NIR 
(constrained by the observed extinction law and albedo) and the 
energy emitted by dust in the MIR-to-submm (constrained by
the observed emissivity spectral energy distribution). The necessity
of the rescaling implies that the high-latitude cirrus emission
is heated by a stronger radiation field than what is measured locally
(if the properties adopted in the dust model are correct). A similar 
conclusion is reached by \citet{YsardA&A2024} when fitting a newer
version of the THEMIS model for diffuse dust: they found that a better match 
between the model output and the observed
dust emissivity is obtained when the MMP LISRF is scaled by 
a factor\footnote{
\citet{YsardA&A2024} and the DustEM code call this factor $G_0$. Usually, 
the name $G_0$ is adopted for the ratio between the total energy radiated 
over the photon energy range [6,13.6] eV (i.e.\ [0.0912,0.2] $\mu$m) and 
the same value from the theoretical estimate of the ISRF in the 'typical' 
ISM made by \citeauthor{HabingBAIN1968} (\citeyear{HabingBAIN1968}; see also, \citealt{TielensApJ1985}).
In units of the Habing flux, it is $G_0=1.14$ for the MMP \citep{DraineBook2011}
and $G_0=0.8$ for the current LISRF, accounting for its lower UV output. For simplicity,
DustEM sets $G_0=1$ for the MMP. Also, the shape of the rescaled spectrum is kept 
fixed, and thus $G_0$ and $U$ are equivalent 
\citep[see also][]{CompiegneA&A2011}.
}
$U=1.4$.

Even without using the models of \citet{HensleyApJ2023} and \citet{YsardA&A2024}, 
the impact of the new LISRF can still be tested, following their approach.
Let's assume for a while that the observed dust emissivity is
the same as that derived from the THEMIS version I use,
when the heating is provided by the MMP LISRF multiplied by a 
factor $U=1.4$ (Fig.~\ref{fig:dustem}, top-right panel);
and that the total energy absorbed from the rescaled MMP LISRF
matches the energy emitted by dust (i.e. the integral of the red 
curve in the bottom-right panel of Fig.~\ref{fig:dustem}, for this 
test). When I use the LISRF derived in this work, I need to multiply
it by a factor $U\approx 1.3$ in order to match the same absorbed
energy: the factor $U$ is slightly smaller than for MMP, because, 
as we have seen, slightly more energy is absorbed by the unscaled 
(i.e. $U=1$) new LISRF. I applied the same procedure to the lower and upper 
boundaries of the uncertainty estimate, by multiplying the former by $U=1.5$ 
and the latter by $U=1.2$: being all spectra normalized to the same energy values, the reduced
shaded area in the top right panel now reflects  only the different uncertainties 
estimated across the spectrum (larger in UV than in NIR).
The corresponding emissivity (and uncertainty) is shown as a green
line in the bottom-right panel of Fig.~\ref{fig:dustem}: while the total absorbed energy is the same by construction, the emissivity computed
using the spectral shape of the new LISRF is sligthly higher in the 
FIR/submm (by 2\% at the peak and tending to 1\% for the summ)
and considerably smaller at shorter wavelengths (by 15\% at 30 $\mu$m
and $\approx$30\% within the MIR bands). Again, as for the output of the
unrescaled model (bottom-left panel) I could match the "observed" 
emissivity (the red curve, for this test), by increasing 
in the dust model the contribution of small grains relatively to large grains. For the \citet{DraineBook2011}
LISRF, $U\approx1.2$ is needed: the emissivity is 4\% higher at the peak
(approaching 1\% in the submm), 5\% lower at 30 $\mu$m and $\approx$12\% lower in the MIR.

\citet{HensleyApJ2023} and \citet{YsardA&A2024} require their models to match the observational constraints within a maximum discrepancy of 20\%. Without changing any of the properties in the dust model, the use of the LISRF derived in this work can result in differences in the predicted dust emissivity as high as 30\%. Therefore, the impact of the new LISRF should be considered in dust models, in particular for the component of small grains responsible for the continuum and band emission in the MIR.

\section{Summary}
\label{sect:summary}

I reevaluated the spectrum of the LISRF from the UV to the NIR, using {\em Gaia} DR3 photometry and stellar atmosphere parameters. In shorts:

\begin{itemize}
\item the contribution of direct starlight to the LISRF has been estimated in the three {\em Gaia} photometric bands, by summing up the fluxes of all stars in the DR3 catalog, and those of nearby, bright, objects in the {\em Hipparcos} catalog  not included in DR3.

\item The full UV-to-NIR spectrum for direct starlight has been computed, using the stellar atmosphere models of \citet{CastelliProc2003} for {\em Gaia} stars with APs, a color-based extrapolation for {\em Gaia} stars without APs, and an additional component for the {\em Hipparcos} stars not in DR3.

\item The full LISRF including direct and diffuse starlight has been derived in the optical, using  {\em Pioneer} 10/11 observations, and in the NIR, using DIRBE maps. After subtracting the contribution of direct starlight obtained from the {\em Gaia}-based spectrum, the maps of the residuals have been used (together with GALEX background maps) as a reference set for the diffuse starlight component of the LISRF.

\item Correlations found in the literature between diffuse starlight (scattered by dust in the ISM) and IRAS 100 $\mu$m emission (from dust at thermal equilibrium)
have been rescaled to match the levels in the reference set and used to derive the diffuse starlight spectrum of the LISRF from the UV to the NIR.

\item Finally, the contribution of both direct and diffuse starlight has been summed to obtain the full LISRF spectrum.
\end{itemize}

The main results of this work are:
\begin{enumerate}
\item the new LISRF spectrum is redder than that of the standard MMP model of \citet{MathisA&A1983}, being $\sim$20 lower in the  $FUV$ and $>30\%$ higher 
from the optical to the NIR. Also, the total energy emitted over the UV-to-NIR spetrum is 
$\sim$30\% higher than the corresponding MMP value. In the optical/NIR, differences with respect 
to the  \citet{DraineBook2011} update of MMP are smaller, the new estimate being at most
$\sim 15\%$ lower (at $\lambda\approx 1\,\mu$m).

\item Because of the different spectral shape, dust models exposed to the new LISRF can 
have MIR emissivities by up to 30\% lower than in the standard case when the MMP is used as a heating field. This difference could lead to significant changes in the constraints of the
population of small grains and of the carriers of the MIR bands.

\item On average, diffuse radiation scattered by dust in the optical contributes to 25\% of $I_\lambda$ for $|b|<50^\circ$. Apparently, the global correlation between dust scattered radiation and dust emission at 100 $\mu$m is stronger than what found in the literature mostly at high Galactic latitude, by a factor $\sim$3 in the optical/NIR.
\end{enumerate}

For the next generation of dust grain models, I recommend the usage of the new LISRF presented in this work (and its comparison with MMP and the lesser used 
formulation by \citealt{DraineBook2011}). 

{\em Upon publication of the paper} I provide online at the CDS: the $J_\lambda$ spectrum (total and separate  direct and diffuse components); the $I_\lambda$ spectral cubes; the {\em Pioneer} $I_\lambda$ maps including the contribution of bright stars.

\begin{acknowledgements}
I am greatly indebted to Elena Pancino for her exquisite guidance through the details of {\em Gaia} and its data products. I also thank L. Magrini and S. Zibetti for many helpful suggestions on the modelling of stellar spectra, Y. Matsuoka for details on the correlations between scattered starlight and infrared radiation, and B. Draine for helpful discussions. 
\\
This work has made use of data from the European Space Agency (ESA) mission
{\it Gaia} (\url{https://www.cosmos.esa.int/gaia}), processed by the {\it Gaia}
Data Processing and Analysis Consortium (DPAC,
\url{https://www.cosmos.esa.int/web/gaia/dpac/consortium}). Funding for the DPAC
has been provided by national institutions, in particular the institutions
participating in the {\it Gaia} Multilateral Agreement.
Part of this work is based on archival data, software or online services provided by the Space Science Data Center, a facility maintained by the Italian Space Agency.
This research has made use of the SIMBAD database and of the
SIMBAD catalogue access tool,
operated at CDS, Strasbourg, France.
I acknowledge the use of data provided by the Centre d'Analyse de Données Etendues (CADE), a service of IRAP-UPS/CNRS (http://cade.irap.omp.eu, \citealt{ParadisA&A2012}). 
I acknowledge the use of the Legacy Archive for Microwave Background Data Analysis (LAMBDA), part of the High Energy Astrophysics Science Archive Center (HEASARC). HEASARC/LAMBDA is a service of the Astrophysics Science Division at the NASA Goddard Space Flight Center.
 \\
For the analysys and visualization of the results, I used
Python \citep{python2009}
and its packages
astropy \citep{astropy2018};
dust\_extinction \citep{dust_extinction};
healpy \citep{ZoncaOJ2019};
matplotlib \citep{matplotlib2007};
numpy \citep{numpy2020};
pandas \citep{pandas2010};
pysynphot \citep{pysynphot2013}.
\end{acknowledgements}

\bibliographystyle{aa} 
\bibliography{DUST} 

\begin{appendix}
    
\section{New LISRF determinations}
\label{app:newe}

\begin{figure}[b]
\centering
\includegraphics[width=\hsize]{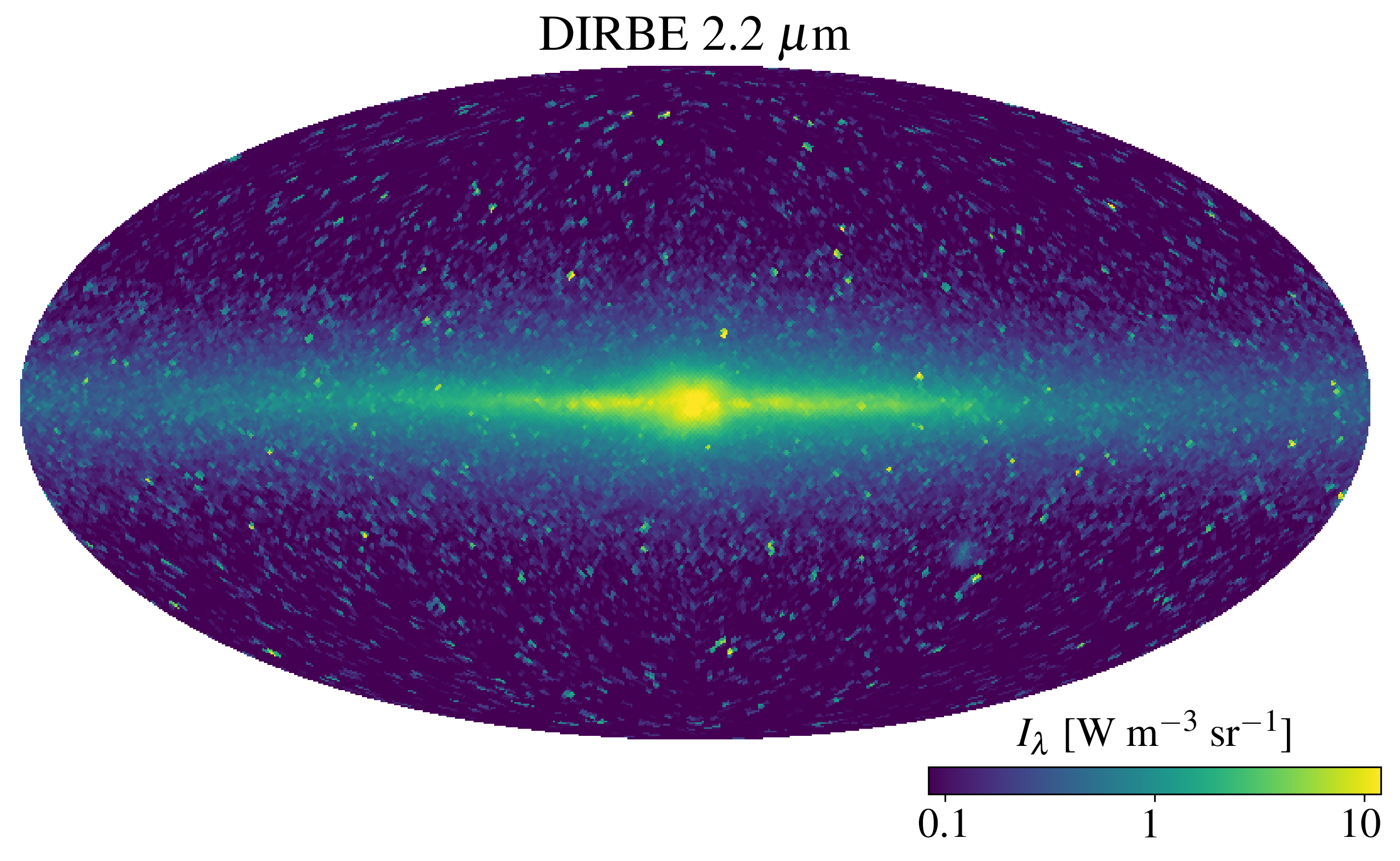}
\caption{
$I_\lambda$ at 2.2 $\mu$m from DIRBE.
}
\label{fig:I_D1}
\end{figure}

I describe here new estimates of $J_\lambda$ in the NIR (obtained from COBE-DIRBE data) and in the optical (from {\em Pioneer}-IPPs data). 

\subsection{COBE-DIRBE}
The DIRBE instrument aboard the COBE satellite observed the whole sky in 10 bands between the
NIR (from 1.25~$\mu$m) and the FIR (to 240~$\mu$m; \citealt{HauserApJ1998}). In this work,
I use the Zodi-Subtracted Mission Average (ZSMA) maps at 1.25, 2.2, 3.5~$\mu$m \citep{ArendtApJ1998}, available in HEALPix format at the “Centre d'Analyse de Donn\'ees Etendues” (CADE)\footnote{http://cade.irap.omp.eu}. The maps were color corrected from the original calibration convention, assuming a blackbody spectrum with $T=3000K$ (i.e.\ the NIR dilute blackbody in MMP; see Sect.~\ref{sect:compa}) 
and the DIRBE color-correction software available at the Legacy Archive for Microwave Background Data Analysis (LAMBDA)\footnote{https://lambda.gsfc.nasa.gov/}.

The DIRBE $I_\lambda$ map at 2.2 $\mu$m is shown in Fig.~\ref{fig:I_D1}.
$J_\lambda$ for the three DIRBE bands considered here is listed in Table~\ref{tab:J_l_new}.
The errors are dominated by the calibration uncertainty (3.1\%; \citealt{HauserApJ1998}).

\label{app:dirbe}
\subsection{Pioneer-IPPs}
\label{app:pioneer}

The {\em Pioneer 10} and {\em 11} probes were equipped with two identical instruments, the Imaging Photopolarimeters  (IPPs; 
\citealt{WeinbergProc1973}), observing 
in a blue band ($B_\mathrm{IPP}$, $\lambda=$ 0.437 $\mu$m) and a red band
($R_\mathrm{IPP}$, $\lambda=$ 0.644 $\mu$m).
On a few occasions, the IPPs mapped a considerable part of the whole sky, while the probes were spinning around their cruise directions: 
this resulted in complex observation patterns, with the field of view of the various pointings (FOVs) ranging from 7.5 to 18 degree$^2$. 
Furthermore, the patterns do not overlap, since the probes 
gradually changed cruise directions \citep{TollerA&A1987, 
GordonApJ1998,LeinertA&AS1998}.

\begin{table}[t]
\caption{New LISRF determinations.}
\label{tab:J_l_new}
\centering
\begin{tabular}{lccc}
                &\multicolumn{3}{c}{$J_\lambda$ [W~m$^{-3}$ sr$^{-1}$]}\\\\
                &1.25~$\mu$m        &2.2~$\mu$m        &3.5~$\mu$m \\\\
         DIRBE  &   0.96$\pm$0.03  &0.32$\pm$0.01    & 0.080$\pm$0.0025   \\\\
                &$B_\mathrm{IPP}$       &$R_\mathrm{IPP}$     &\\
                &0.437~$\mu$m        &0.644~$\mu$m               &\\ \\
$V\le 6.5$ stars      & 0.34$\pm$0.04 & 0.25$\pm$0.03 &\\
{\em Pioneer 10/11}& 1.17$\pm$0.20 & 1.41$\pm$0.15 &\\
Total&   1.51$\pm$0.21  &1.66$\pm$0.15     &\\
\end{tabular}
\end{table}

\begin{figure*}
\centering
\includegraphics[width=\hsize]{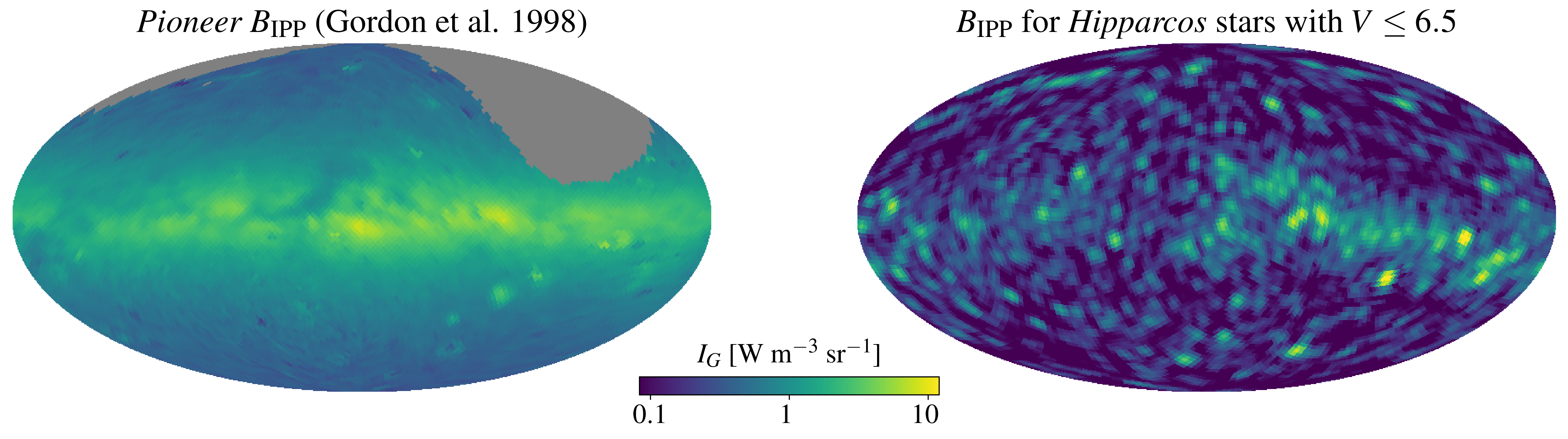}
\caption{
Left: {\em Pioneer} $B_\mathrm{IPP}$ map from \citet{GordonApJ1998}. Right: reconstructed contribution from $V \le 6.5$ 
stars using the {\em Hipparcos} catalog, smoothed
with a FWHM=4$^\circ$ Gaussian (to  match 
approximately the resolution of IPP maps).
}
\label{fig:I_BIPP}
\end{figure*}

The data was distributed after removing the contribution of the 
brighter stars, estimated using the entries in  the Yale Bright Stars Catalog and in the USNO
Photoelectric Catalog (for additional stars with $V<8$): the
choice is equivalent to select all stars with $V<6.5$
\citep{TollerA&A1987}. In order to recover this contribution,
I used the {\em Hipparcos} catalog to select $V<6.5$ stars; they are $\sim$8800, while the original list subtracted 
from the IPP maps had $\sim$12500 entries \citep{LeinertA&AS1998}. For these stars,
I converted the {\em Hipparcos/Tycho} 
$B_\mathrm{T}$ and $V_\mathrm{T}$ magnitudes
from the catalog to those in the IPP system (with the 
transformations of \citealt{MatsuokaApJ2011}).
Finally, fluxes in physical units were obtained using the 
zero points from \citet{GordonApJ1998} and summed
over a HEALPix grid. The reconstructed
map of $I_\lambda$ for the bright stars is shown in
Fig.~\ref{fig:I_BIPP} for the $B_\mathrm{IPP}$ band;
$J_\lambda$ values are listed in Table~\ref{tab:J_l_new}.
For an additional check, I downloaded more recent versions
of the Yale Bright Stars and Photometric catalogs
(\citealt{HoffleitBook1991} and \citealt{MermilliodA&A1987}, respectively) from 
SIMBAD\footnote{https://SIMBAD.cds.unistra.fr/}.
After applying the $V<8$ cut, the merged catalog contains $\sim17000$ stars.
I converted the tabulated Johnson $B$ and $V$ magnitudes to those in the 
{\em Hipparcos/Tycho} system (using the transformations in
\citealt{PancinoA&A2022}); and then proceeded as above
for the {\em Hipparcos} catalog. 
The difference with the previous estimate
is $\la15\%$; I adopt it as the uncertainty in the
estimate of $J_\lambda$, accounting for the  
magnitude transformations and the differences between the 
stellar catalogs I used and the original catalog  used for star removal.

\citet{GordonApJ1998} produced maps from eleven IPP observations 
taken in 1973-74, when the {\em Pioneer}'s were sufficiently distant 
from the Sun and the contamination from interplanetary dust negligible. Because of the subtraction of bright stars, the IPP maps should include the total radiation from all stars with $V>6.5$, plus the diffuse radiation scattered by dust (or any other source of radiation 
beyond stars, such as the dust-related extended red emission
investigated by \citealt{GordonApJ1998}).
I converted the \citet{GordonApJ1998} 
maps\footnote{I used the {\em $1^\mathrm{st}$ iteration}
maps and the pre-processed information on the photometry and position of each FOV, available at 
\url{https://karllark.github.io/data_pioneer10_11_ipp.html}
.}
to the current format to derive $I_\lambda$
(shown in Fig.~\ref{fig:I_BIPP} for  $B_\mathrm{IPP}$); and
$J_\lambda$, after filling the regions unobserved by the IPPs with 
the average $I_\lambda$ trend with latitude 
(Table~\ref{tab:J_l_new}). Before producing the maps,
\citet{GordonApJ1998} filtered the original FOV data 
to remove observations close to the Sun and fluxes corrupted during data transmission; but also data with "incorrect subtraction of bright stars".
In order to check the importance of the incorrect star subtraction in the original dataset, 
I produced new maps directly from the FOV list provided by the authors, without applying filtering. Curiously, in the unfiltered maps 
many bright stars are visible, even as bright as $V\le2$. Some of them can still be seen, as residuals, in the maps of \citet{GordonApJ1998}: for example,
when comparing the two maps in Fig.~\ref{fig:I_BIPP}, one can identify
easily $\alpha$ Boo (Arcturus, near the central meridian towards the Galactic North Pole) and $\alpha$ Car (Canopus, below the Galactic plane on the top-right of the Large Magellanic Cloud).
From the unfiltered maps, I can derive $J_\lambda$ values higher 
than those from the maps of \citet{GordonApJ1998}, by
15\% and 7\% in the $B_\mathrm{IPP}$ and $R_\mathrm{IPP}$ bands, respectively.
The difference between the unfiltered {\em Pioneer} maps, 
which still apparently includes bright stars, and the maps of 
\citet{GordonApJ1998}, where most (thought not all) the flux from 
bright stars is removed, are used as a conservative estimates of the uncertainties on IPP estimates of $J_\lambda$, to which I added a further 8\% calibration uncertainty
\citep{HannerProc1976}. The adopted error on the IPP-based $J_\lambda$ is given in Table~\ref{tab:J_l_new}.

Finally, I sum the contribution of the bright stars and of the
{IPP} maps to derive the total estimates of $J_\lambda$ 
(Table~\ref{tab:J_l_new}).

\section{Validation of synthetic spectra}
\label{app:xm}

The reliability of the synthetic spectra obtained in Sect.~\ref{sect:spectrum}
is controlled by deriving $J_\lambda$ for the cross-matches between the 
{\em Gaia}/{\em Hipparcos} catalogs and other independent photometric catalogs:
the value obtained by the summation of the stellar fluxes in each photometric band 
of the independent catalog is compared with the synthetic photometry derived 
over the same band from the $J_\lambda$ spectrum. I use the TD-1 catalog 
\citep{ThompsonBook1978} in the UV, the 2MASS PointSource Catalog \citep{SkrutskieAJ2006}
in the NIR, for {\em Gaia} and {\em Hipparcos}-derived spectra,
and $BVJHK$ photometry from SIMBAD, for {\em Hipparcos} only.

\begin{figure*}
\centering
\includegraphics[width=\hsize]{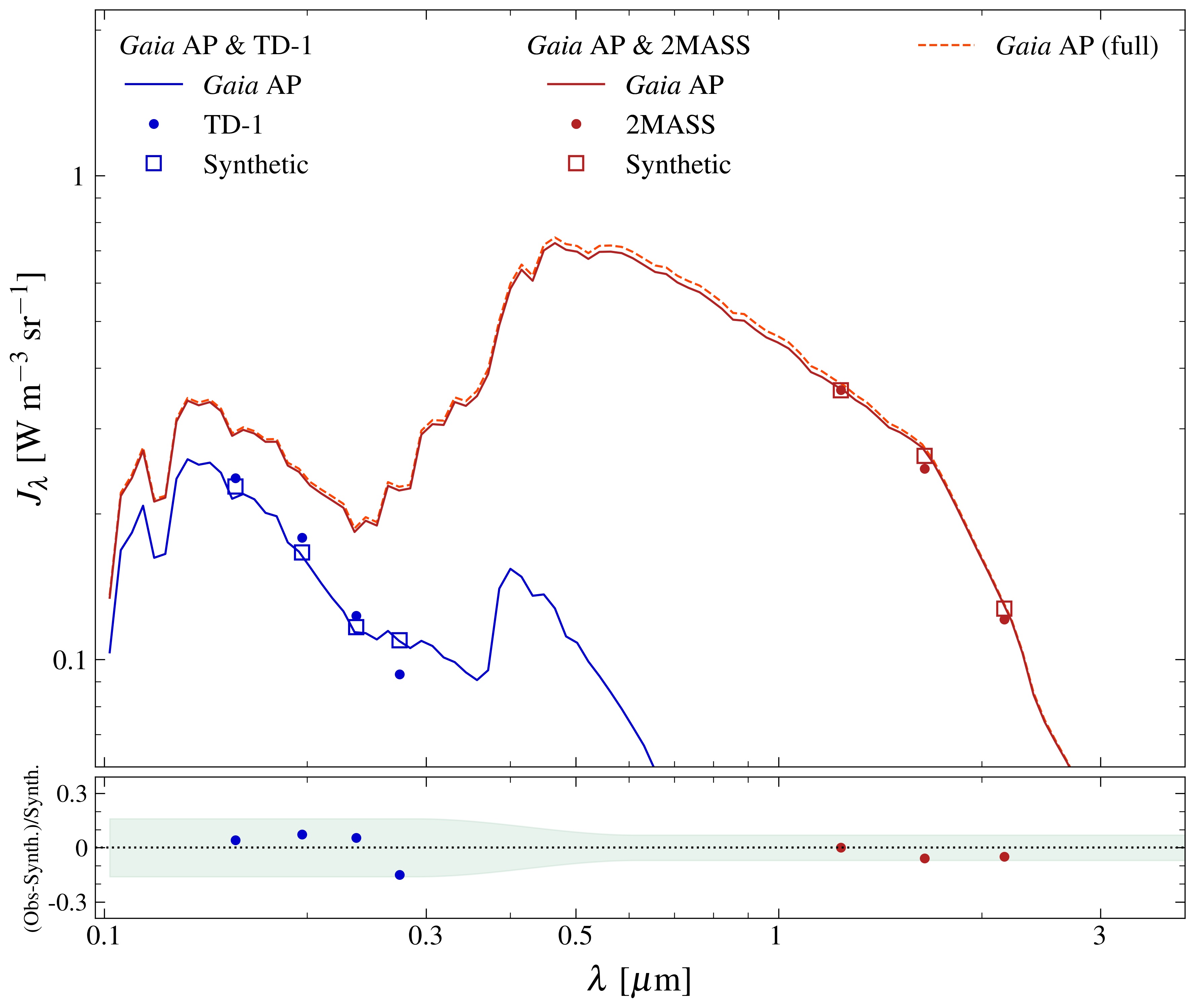}
\caption{
$J_\lambda$ for the cross-matches between {\em Gaia} AP stars and the TD-1 and 2MASS catalogs.
I show the spectrum, synthetic photometry and observation for each crossmatch. I also show
for comparison the {\em Gaia} AP spectrum.
The relative difference between observations and synthetic photometry is shown in the bottom panel.
The shaded area highlights the uncertainty I adopt for the estimate.
}
\label{fig:LISRF_XM_Gaia}
\end{figure*}

\subsection{Gaia}

Using SIMBAD, I searched the {\em Gaia} DR3 identifier for each star in the TD-1 catalog: 
out of the $\sim$31000 stars in TD-1, I found a counterpart in {\em Gaia} for $\sim$29500; 
the selection further restricts to $\sim$19000 stars if only {\em Gaia} AP objects 
are considered
for which the spectrum can be estimated directly (see Sect.~\ref{sect:gaia_spectrum}).
The $J_\lambda$ spectrum for this selection, together with synthetic and observed 
datapoints, is shown in Fig.~\ref{fig:LISRF_XM_Gaia}. Judging on the difference, I adopt
for the spectrum a relative error of 16\% in this range. Even though the spectrum for
this selection is on overage only 40\% of the full {\em Gaia} (AP + non-AP extrapolation) spectrum,
I assume that the same relative error can be applied in this range for the latter.

The DR3 cross-match between 2MASS and {\em Gaia} contains
 $\sim$330 million {\em best-neighbour} counterparts to {\em Gaia} AP stars.
Again the total spectrum, the synthetic estimate and the result from flux summation are shown in Fig.~\ref{fig:LISRF_XM_Gaia}. In this range, the difference reduces to 7\%. The total spectrum 
for this selection is very close to the full {\em Gaia AP} spectrum across all wavelengths
(also shown in the Figure); in the NIR, it is 40\% of the full {\em Gaia} (AP + non-AP extrapolation) spectrum.

\begin{figure*}
\centering
\includegraphics[width=\hsize]{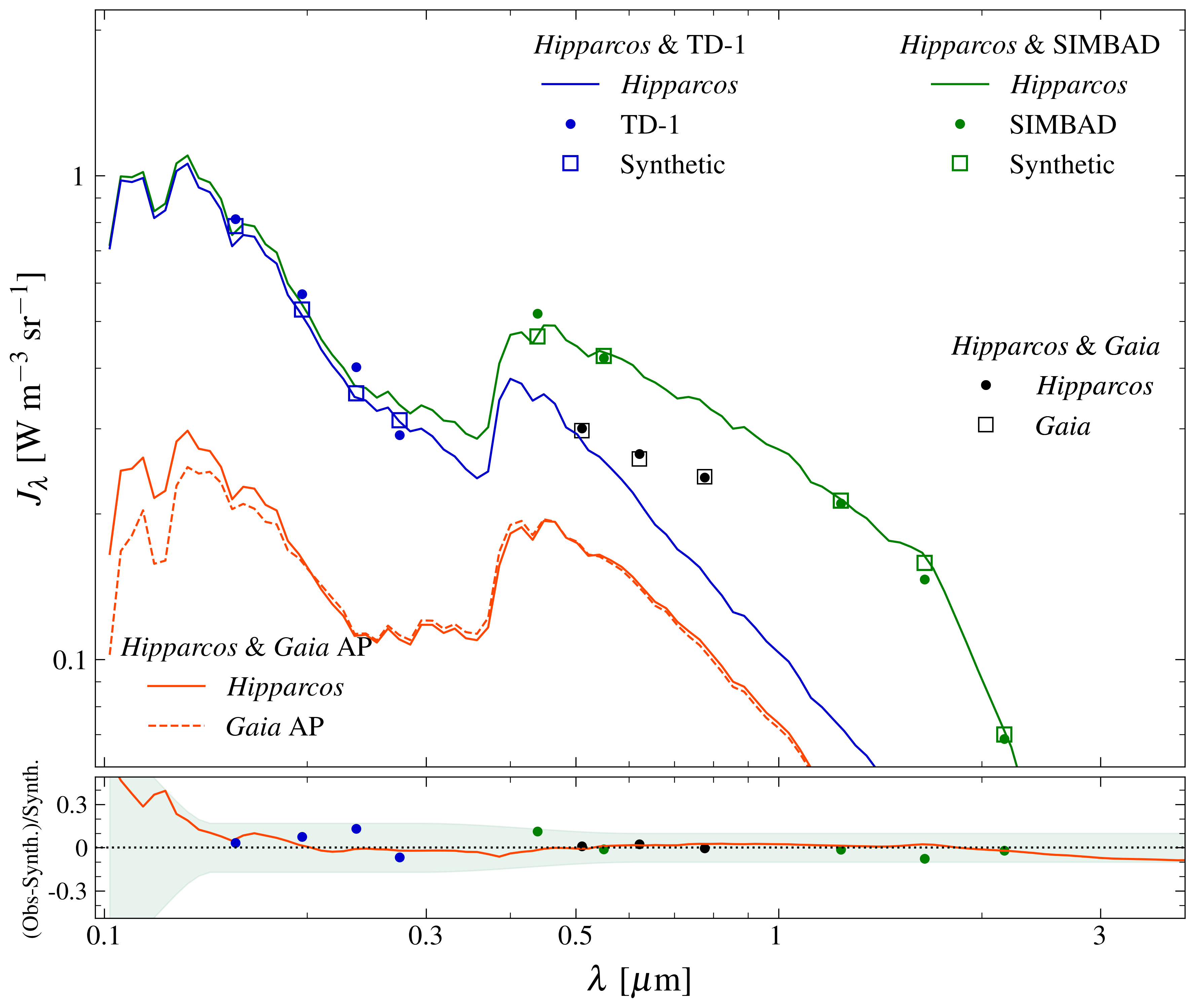}
\caption{
Same as Fig.~\ref{fig:LISRF_XM_Gaia}, but for the cross-matches between {\em Hipparcos} and the 
TD1 catalog and a SIMBAD photometry search. I also show $J_\lambda$ in the {\em Gaia} bands
derived independently from {\em Hipparcos} and {\em Gaia}, for the official crossmatch between 
the two catalogs; and the independent spectra estimated from matched {\em Hipparcos} and {\em Gaia}
AP stars. For the {\em Hipparcos}/{\em Gaia} comparison in the lower panel, the reference values 
(i.e.~{\em Synth.}) are the {\em Gaia} photometry and spectrum.
}
\label{fig:LISRF_XM_Hipparcos}
\end{figure*}

\subsection{Hipparcos}

According to SIMBAD, there are $\sim$23500 TD-1 stars with an {\em Hipparcos} identifier.
From the difference between synthetic and observed datapoints in Fig.~\ref{fig:LISRF_XM_Hipparcos}, 
I adopt for the spectrum a relative error of 17\% in this range. In this case,
the spectrum for the selection is 2.5$\times$ higher than the one I derive in Sect.~\ref{sect:hipparcos_spectrum}
for the {\em Hipparcos} stars not included in the {\em Gaia} catalog, which are only $\sim$1500.

Using SIMBAD, I found a complete set of $BVJHK$ photometry for $\sim$94\% of the {\em Hipparcos} stars. Most of the recovered photometry comes from the Tycho-2 catalog \citep[for $B$ and $V$]{HoegA&A2000} and the 2MASS PointSource Catalog \citep[for $J$, $H$ and $K$]{SkrutskieAJ2006}.
The comparison between synthetic and observed datapoints
(see Fig.~\ref{fig:LISRF_XM_Hipparcos}) suggests an overall uncertainty of 
the spectrum of 10\% over the optical/NIR range. In this case, the spectrum is $\sim6\times$
higher than for the selection used in Sect.~\ref{sect:hipparcos_spectrum}

\subsection{Gaia vs Hipparcos}
\label{sect:gvsh}

I also checked the accuracy in the derivation of {\em Gaia}-band  photometry
from the original {\em Hipparcos} photometry (Sect.~\ref{sect:hipparcos}). In Fig.~\ref{fig:LISRF_XM_Hipparcos} I show the comparison between $J_\lambda$ in the {\em Gaia} bands, obtained from 
{\em Hipparcos} fluxes and directly from {\em Gaia} fluxes, for the official {\em best-neighbour}
crossmatch between the two catalogs ($\sim$100000 stars). The relative difference is within 2\% (See also Sect.~\ref{sect:hipparcos} and Table~\ref{tab:J_l}).

\begin{figure}
\centering
\includegraphics[width=\hsize]{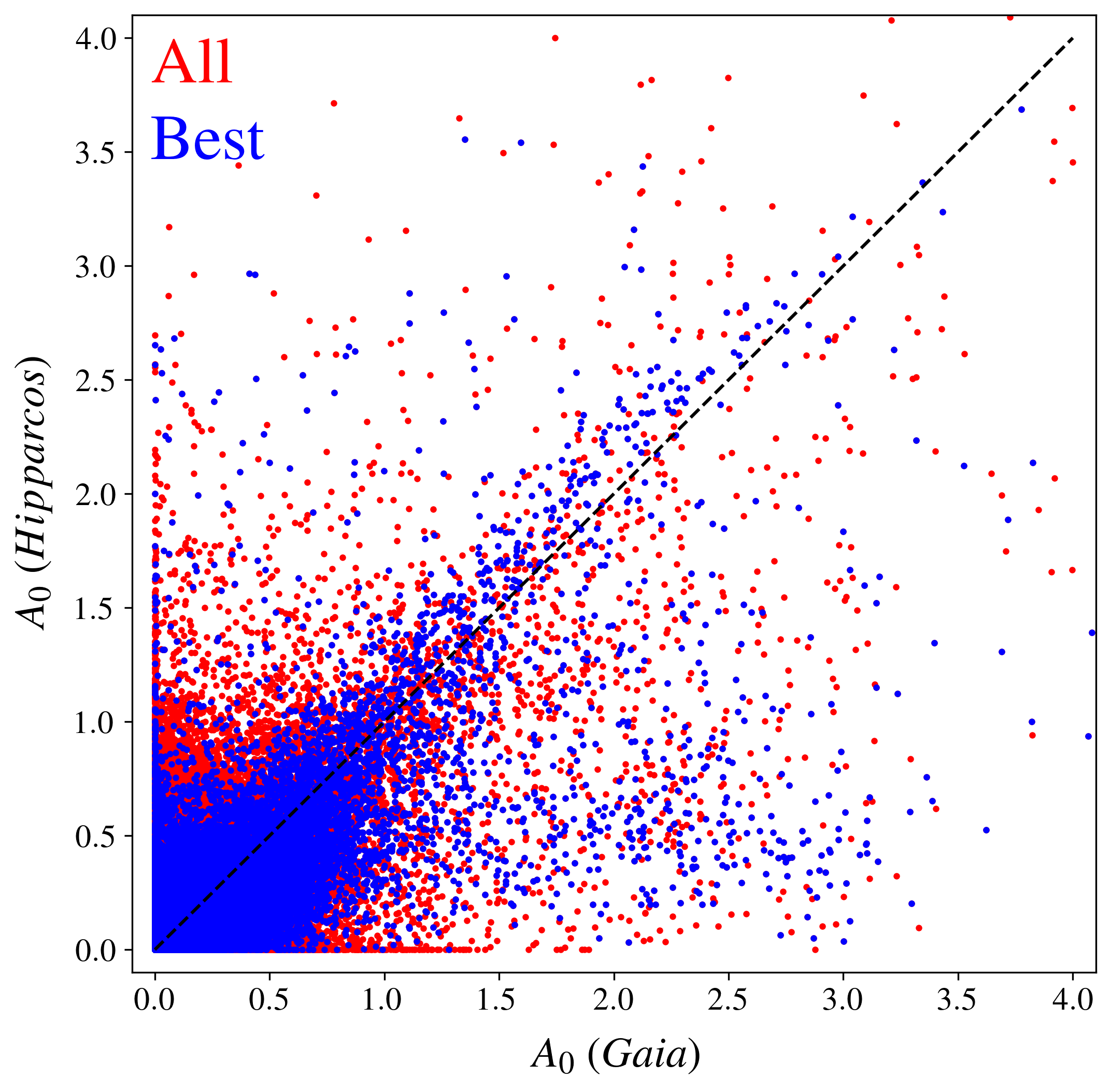}
\caption{
Comparison between $A_0$ values from the {\em Gaia} AP dataset and those derived from independent stellar parameters from the SIMBAD database and the {\em Hipparcos} catalog. Red datapoints show all stars in the {\em Gaia-Hipparcos} crossmatch, blue ones those with the best independent dataset (all stellar parameters recovered and extinction derived from six colors; see Sect.~\ref{sect:hipparcos_spectrum} for details).
The dashed line shows the 1-1 correspondence.
}
\label{fig:A0_comp}
\end{figure}

In analogy, I checked the total spectra derived independently from the two ways described in Sect.~\ref{sect:spectrum}, for a subsample restricted to the stars with {\em Gaia} APs
($\sim$65000 stars). 
The crossmatch gives the opportunity of comparing the values of $A_0$ from {\em Gaia}, derived self-consistently with the other APs, and those computed from the 
optical-NIR photometry available for {\em Hipparcos} stars, after assuming the parameters for stellar atmospheres derived from
SIMBAD and the {\em Hipparcos} catalog (see Sect.~\ref{sect:hipparcos_spectrum}). The comparison is shown in Fig.~\ref{fig:A0_comp}: there is a general agreement between the two estimates, but with a large
scatter, even for the best case in which all stellar parameters of {\em Hipparcos} stars are retrieved 
from the independent dataset, and $A_0$
is derived from the average of six colors.
The scatter is probably due to the
extinction-temperature degeneracy: if the predicted 
intrinsic stellar spectrum is, e.g., too blue, the corresponding $A_0$ is biased towards higher values 
in order to compensate and match the observed photometry.
While this issue affects also AP estimates \citep{AndraeA&A2023}, it is likely more 
important in the independent, non self-consistent, estimates based on SIMBAD and the {\em Hipparcos} catalog.
Nevertheless, the sum over the whole cross-match catalog results in total spectra (also shown in Fig.~\ref{fig:LISRF_XM_Hipparcos}) which are still 
within the uncertainties already assumed 
in this section. Averaging over a large number of objects
(most of which have small $A_0$ values because of their closeness) reduces the effects of the scatter in $A_0$ 
(and of the differences between {\em Gaia} APs and the independent stellar parameters). Only for 
$\lambda\le 0.11 \mu$m the uncertainties are larger, probably because of the issues discussed so-far, but also  
because of the different gridding in the table of model spectra used for the {\em Hipparcos} estimate. 
Attaching more confidence to
the {\em Gaia}-derived spectra, I raise the uncertainty for {\em Hipparcos} spectrum in this
range accordingly.

\end{appendix}

\end{document}